\documentclass[journal]{IEEEtran}
\usepackage{amsmath,amsfonts}
\usepackage{algorithmic}
\usepackage{algorithm}
\usepackage{array}
\usepackage[caption=false,font=normalsize,labelfont=sf,textfont=sf]{subfig}
\usepackage{textcomp}
\usepackage{stfloats}
\usepackage{url}
\usepackage{verbatim}
\usepackage{graphicx}
\usepackage{cite}
\usepackage{hyperref}
\usepackage{cleveref}
\usepackage{silence}
\usepackage{microtype}
\graphicspath{{figs/}{figures/}{pictures/}{images/}{./}}

\usepackage{booktabs}
\usepackage{mathptmx}
\usepackage{ccicons}
\usepackage{csquotes}

\newcommand{\ST}{ST }

\newcommand{\etal}{et al.\ }

\newcommand{\taxaxisformat}[1]{\emph{#1}}
\newcommand{\taxdimformat}[1]{\textit{#1}}
\newcommand{\taxcodeformat}[1]{\textsc{#1}}
\newcommand{\taxtaskformat}[1]{\textsf{#1}}

\newcommand{\deftaxaxis}[2]{\expandafter\def\csname taxaxisname@#1\endcsname{#2}}
\newcommand{\deftaxdim}[2]{\expandafter\def\csname taxdimname@#1\endcsname{#2}}
\newcommand{\deftaxcode}[2]{\expandafter\def\csname taxcodename@#1\endcsname{#2}}
\newcommand{\deftaxtask}[2]{\expandafter\def\csname taxtaskname@#1\endcsname{#2}}

\newcommand{\taxaxis}[1]{\taxaxisformat{\csname taxaxisname@#1\endcsname}}
\newcommand{\taxdim}[1]{\taxdimformat{\csname taxdimname@#1\endcsname}}
\newcommand{\taxcode}[1]{\taxcodeformat{\csname taxcodename@#1\endcsname}}
\newcommand{\taxcodeplural}[1]{\taxcodeformat{\csname taxcodename@#1\endcsname s}}
\newcommand{\taxtask}[1]{\taxtaskformat{\csname taxtaskname@#1\endcsname}}

\deftaxaxis{framework}{What--Why--How}
\deftaxaxis{what}{\textbf{What}}
\deftaxaxis{why}{\textbf{Why}}
\deftaxaxis{how}{\textbf{How}}
\deftaxaxis{data}{data}
\deftaxaxis{task}{task}
\deftaxaxis{visualization}{visualization}

\deftaxdim{modality}{modality}
\deftaxdim{resolution}{resolution of observation}
\deftaxdim{condition}{condition}
\deftaxdim{metadata}{metadata}
\deftaxdim{data-components}{data components}
\deftaxdim{visualized-elements}{visualized elements}
\deftaxdim{biological-scale}{biological scale}
\deftaxdim{analytical-domain}{general analytical domain}
\deftaxdim{layout}{layout}
\deftaxdim{abstraction}{abstraction}
\deftaxdim{comparison}{comparative design}
\deftaxdim{scalability}{scalability strategy}
\deftaxdim{contextualization}{contextualization and communicative design}
\deftaxdim{chart-type}{chart type}
\deftaxdim{interaction-intent}{interaction intent}

\deftaxtask{gene-expression}{gene expression mapping}
\deftaxtask{svg-discovery}{spatially variable gene (SVG) discovery}
\deftaxtask{coexpression}{co-expression/module analysis}
\deftaxtask{pathway-enrichment}{functional pathway enrichment}
\deftaxtask{cell-type}{cell-type mapping}
\deftaxtask{domain-discovery}{spatial domain discovery}
\deftaxtask{tissue-structure}{tissue structure}
\deftaxtask{cell-communication}{cell--cell communication}
\deftaxtask{cellular-niches}{cellular niches}
\deftaxtask{spatial-proximity}{spatial proximity}
\deftaxtask{tissue-boundaries}{tissue boundaries}
\deftaxtask{lineage-trajectory}{lineage \& trajectory inference}
\deftaxtask{evolving-processes}{evolving biological processes}
\deftaxtask{cross-sample}{cross-sample comparison}
\deftaxtask{cross-modal}{cross-modal alignment}
\deftaxtask{deconvolution}{spatial deconvolution}
\deftaxtask{reconstruction-3d}{3D tissue reconstruction}

\deftaxcode{imaging}{imaging-based}
\deftaxcode{sequencing}{sequencing-based}
\deftaxcode{mapped}{deconvolved or computationally mapped}
\deftaxcode{molecular}{molecular}
\deftaxcode{subcellular}{sub-cellular}
\deftaxcode{cellular}{cellular}
\deftaxcode{multicellular}{multi-cellular}
\deftaxcode{tissue-unit}{functional tissue unit}
\deftaxcode{full-fov}{full field-of-view}
\deftaxcode{intercellular}{intercellular}
\deftaxcode{tissue}{tissue}
\deftaxcode{systemic}{systemic or integrative}
\deftaxcode{categorical}{categorical groups}
\deftaxcode{ordered}{ordered variable}
\deftaxcode{discrete}{discrete}
\deftaxcode{continuous}{continuous}
\deftaxcode{neither}{neither}
\deftaxcode{expression}{gene expression values}
\deftaxcode{coordinates}{spatial coordinates}
\deftaxcode{native-2d}{native 2D}
\deftaxcode{true-3d}{true 3D}
\deftaxcode{reconstructed-3d}{3D reconstructed}
\deftaxcode{rna-velocity}{RNA velocity-derived dynamics}
\deftaxcode{temporal}{temporal variables}
\deftaxcode{annotations}{biological annotations}
\deftaxcode{observations}{individual observations}
\deftaxcode{features}{features}
\deftaxcode{relationships}{relationships between entities}
\deftaxcode{aggregates}{aggregate statistics}
\deftaxcode{physical}{physical-space}
\deftaxcode{latent}{latent-space}
\deftaxcode{latent-layout}{latent layout}
\deftaxcode{linear}{linear}
\deftaxcode{circular}{circular}
\deftaxcode{force-directed}{force-directed graph}
\deftaxcode{realistic}{realistic}
\deftaxcode{partial}{partial abstraction}
\deftaxcode{abstract}{complete abstraction}
\deftaxcode{juxtaposition}{juxtaposition}
\deftaxcode{juxtaposed}{juxtaposed}
\deftaxcode{superposition}{superposition}
\deftaxcode{explicit}{explicit encoding}
\deftaxcode{no-comparison}{no comparison}
\deftaxcode{text-graphics}{textual or graphical annotations}
\deftaxcode{highlighting}{visual highlighting}
\deftaxcode{references}{reference structures}
\deftaxcode{item-level}{item-level display}
\deftaxcode{aggregation}{aggregation}
\deftaxcode{sampling}{sampling or filtering}
\deftaxcode{focus-context}{focus-and-context navigation}
\deftaxcode{scatterplot}{scatter plot}
\deftaxcode{heatmap}{heatmap}
\deftaxcode{violin}{violin plot}
\deftaxcode{hexplot}{hexplot}
\deftaxcode{bar-plot}{bar plot}
\deftaxcode{dot-plot}{dot plot}
\deftaxcode{spatial-map}{spatial map}
\deftaxcode{annotation-how}{annotation}

\deftaxcode{histology-image}{histological image}
\deftaxcode{dimensionality-reduction}{dimensionality-reduction plot}
\deftaxcode{chord-diagram}{chord diagram}
\deftaxcode{network-graph}{network graph}
\deftaxcode{line-chart}{line chart}
\deftaxcode{pie-glyph}{pie glyph}
\deftaxcode{select}{select}
\deftaxcode{explore}{explore}
\deftaxcode{reconfigure}{reconfigure}
\deftaxcode{encode}{encode}
\deftaxcode{abstract-elaborate}{abstract/elaborate}
\deftaxcode{filter}{filter}
\deftaxcode{connect}{connect}

\deftaxcode{task-molecular}{molecular \& functional landscapes}
\deftaxcode{task-spatial}{spatial entity \& domain identification}
\deftaxcode{task-relational}{spatial relational networks}
\deftaxcode{task-comparative}{comparative \& multi-dataset integration}
\deftaxcode{task-spatiotemporal}{spatiotemporal dynamics \& perturbations}

\newcommand{\taxaxistext}[1]{\taxaxisformat{#1}}

\begin{document}

\title{How Do We Visualize Space in Molecular Biology? A Study of Spatial
Transcriptomics Visualization Practices
}


\author{
Denisse Chacón-Ramírez,
Mark S. Keller,
Eric Mörth,
Nils Gehlenborg,
Marc Streit,
and Andreas Hinterreiter
\thanks{D. Chacón-Ramírez, M. Streit, and A. Hinterreiter are with
Johannes Kepler University Linz, Linz, Austria.
E-mail: \{denisse.chacon\_ramirez, marc.streit,
andreas.hinterreiter\}@jku.at.}
\thanks{M. S. Keller, E. Mörth, and N. Gehlenborg are with
Harvard Medical School, Boston, MA, USA.
E-mail: \{mark\_keller, nils\}@hms.harvard.edu and
ericmoerth@fas.harvard.edu.}
\thanks{Manuscript received XX, XXXX; revised XX, XXXX.}
}




\maketitle

\begin{abstract}
     A cell's identity depends on where it sits in tissue: for example, a macrophage behaves differently in a tumor core than at its edge. Spatial transcriptomics has transformed how we study this by recovering that lost coordinate, but it does so by producing data that is simultaneously high-dimensional, multimodal, and uncertain. Visualizing this combination is a hard problem in its own right, and one that warrants an assessment of how the field currently represents it, what has worked, and what is still missing. We surveyed 148 papers and 1,824 figure panels using a \emph{What--Why--How} coding framework grounded in Munzner's nested model, connecting the data represented, the biological tasks motivating each visualization, and the design choices through which they are expressed; a subset of the surveyed work also contributed dedicated interactive visualization software that was not necessarily reflected in the static figures, and we looked at what interaction capabilities those tools supported as well. We close by outlining where the field stands and the challenges ahead for bioinformatics and visualization researchers to tackle together.
\end{abstract}







\section{Introduction}

Spatial transcriptomics (ST) reconnects molecular measurements with the tissue structure~\cite{Marx2021}, creating a visualization problem unlike anything in conventional single-cell transcriptomics analysis. Modern omics technologies capture \emph{where} and \emph{when} molecules act within tissues, producing data whose growth in volume and in spatiotemporal resolution strain our capacity for visual representation~\cite{Seydel2025, Nam2024}. A single \ST experiment comprises multiple interdependent layers: gene expression matrices, spatial coordinates, high-resolution tissue images, segmentation masks, and more. Each layer may carry its own uncertainty and operate at a different biological scale~\cite{Velten2023_1, Keller2024, Xia2025}.

Reasoning across these layers relies heavily on visual interpretation, which is why encoding choices in published figures carry particular weight. In computational and methodological papers, published figures are not a final presentation step; they are frequently where spatial biological reasoning is first performed and then codified. The encoding choices made in these figures become the visual conventions through which an entire scientific community interprets and communicates spatial findings. Yet those choices are made largely without guidance from visualization research, and their cumulative effect has never been characterized. To address this gap, we systematically examine how published figures of \ST data connect research questions, data components, and visual encodings.

Prior work relevant to this study falls into two broad categories. The first comprises reviews of \ST technologies, data modalities, and their biological applications, written primarily for computational biology and biology audiences. These reviews catalog and compare experimental platforms, their resolution, throughput trade-offs, and their suitability for specific tissues or diseases~\cite{Moses2022, Williams2022, Rao2021, Jain2024, Park2023, Choe2023, Chen2023, Lim2025, Cheng2023}; these are indispensable for understanding what \ST measures and why it matters, but do not treat visual encoding as a research subject in its own right. The second comprises state-of-the-art reports (STARs) from the visualization community on adjacent biological domains: Pretorius \etal~\cite{Pretorius2017} survey live cell imaging visualization; Beyer \etal~\cite{Beyer2022} survey visualization across the connectomics pipeline; Ehlers \etal\cite{Ehlers2025} survey biological network visualization; and Nusrat \etal~\cite{Nusrat2019} survey genomic data visualization broadly. Each of these reports addresses a domain that shares elements of our problem space: dynamic imaging, spatial embedding, relational structure, or high-dimensional molecular data. Closest to our approach, Keller \etal~\cite{Keller2025} applied a comparable figure-panel coding methodology to single-cell atlas visualization. However, our focus differs in scope: we specifically analyze data that must remain aligned with native spatial position and tissue morphology, making spatial grounding the core axis of our coding framework. Yet no existing STAR characterizes visualization practice for spatially resolved transcriptomic data specifically, or connects visual design to the structure of spatial biological reasoning as we do here.

We present a systematic study of \ST visualization practices organized around a \emph{What--Why--How} coding framework grounded in Munzner's nested model~\cite{munzner2015visualization}. We assembled a corpus of 148 papers and 1,824 qualifying figure panels published between 2018 and 2025, spanning computational methods and visualization tools for spatially resolved data, and separately coded the interaction affordances of \ST visualization systems using the taxonomy of Yi \etal~\cite{Yi2007}. Our findings show that spatial grounding remains prominent in tasks such as cell-type mapping and spatial domain discovery, but recedes in functional, relational, and some comparative analysis, where abstract charts often disconnect findings from their anatomical context. In several of these cases, this recession occurs even where spatial position remains biologically relevant, suggesting a lack of visual idioms rather than a deliberate choice to omit space. Addressing this loss presents an important design opportunity for the visualization community. Toward that end, this article contributes a systematic corpus with an accompanying interactive browser (https://denisseram.github.io/STAR-spatialsc/), the \emph{What--Why--How} coding framework itself, an empirical characterization of visualization and interaction patterns across major reasoning tasks, and a synthesis of open challenges around spatial grounding, uncertainty, scalability, and multiscale reasoning.



\section{Biological Background}

\ST produces data that simultaneously encodes high-dimensional molecular profiles, spatial coordinates, and tissue morphology. Interpreting this data depends simultaneously on molecular, spatial, and anatomical reasoning. This section aims to provide an overview of the field, reviewing the biological phenomena being measured, the technologies that generate these measurements, and the resulting data structures, providing the foundational knowledge needed to interpret the coding framework and results in Sections \ref{sec:taxonomy}--\ref{sec:discussion}.

\subsection{From Molecular Profiling to Spatial Biology}

\begin{figure}[bt]
  \centering
  \includegraphics[width=0.45\textwidth]{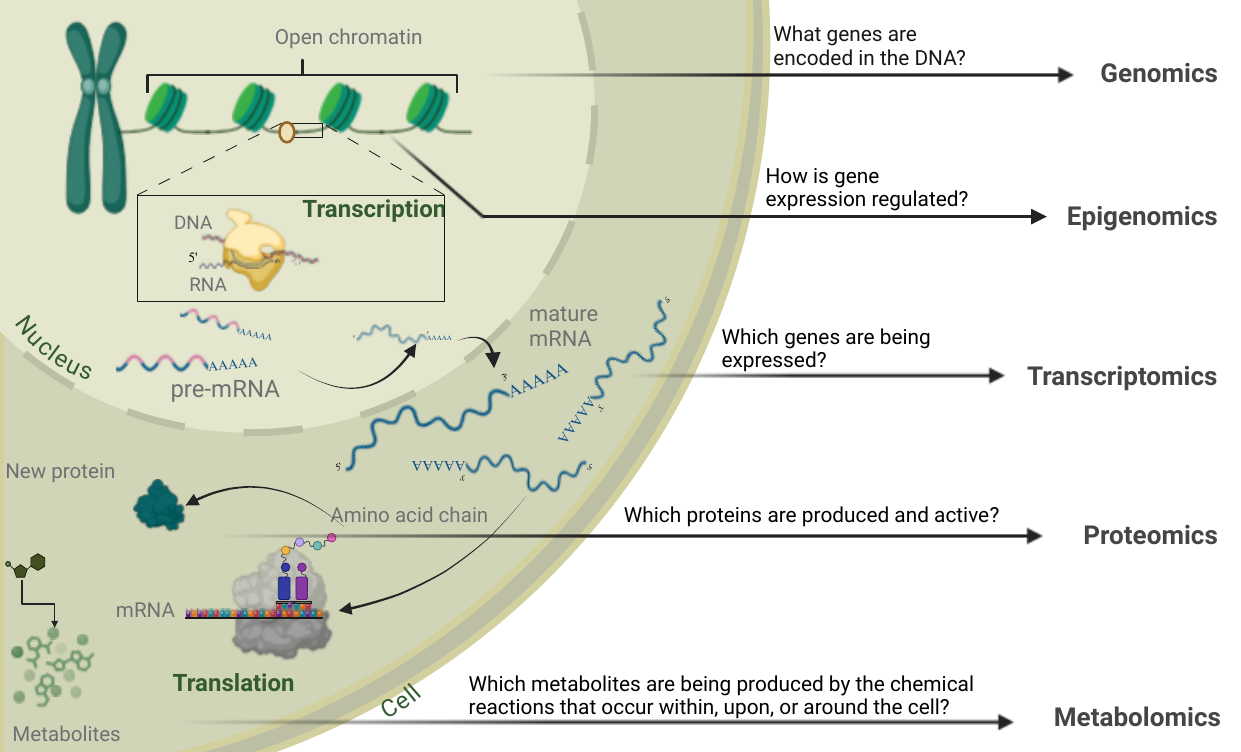}
  \caption{
    The central dogma of molecular biology and associated omics technologies, from genome to transcript to protein to metabolite. Figure created with BioRender.}
    \label{fig:centraldogma}
\end{figure}

Although every cell in a multi-cellular organism carries essentially the same DNA, cells differ in form and function because of \emph{gene expression}: the process by which a cell transcribes genes into messenger RNA (mRNA). The full set of mRNA transcripts, the \emph{transcriptome}, snapshots a cell's functional state, capturing its identity, activity, and environmental response. Measuring this molecular state is a central goal of transcriptomics and is conventionally understood within the framework of the central dogma of molecular biology~\cite{CRICK1970}. Modern molecular biology organizes such measurements into complementary \enquote{omics} layers~\cite{Dai2022} (\Cref{fig:centraldogma}). Among these, transcriptomics is the most widely adopted phenotyping approach because mRNA is both information-rich and highly accessible through high-throughput sequencing~\cite{Swift2021}.

Transcriptomics has progressively recovered finer biological detail: bulk RNA sequencing averages expression across entire tissues, collapsing cellular diversity into a single profile, while single-cell RNA sequencing (scRNA-seq) resolves individual cells and reveals rare types, transient states, and differentiation trajectories that bulk measurements obscure~\cite{Macosko2015, Hwang2018}. Yet, scRNA-seq requires dissociating tissue into single-cell suspensions, which destroys the spatial relationships that organized those cells in the tissue~\cite{Heumos2023}.

This loss of spatial context creates a gap in our understanding of tissue architecture because spatial organization is itself a carrier of biological information. Cells do not function in isolation; their behavior is shaped by their neighbors, their position within anatomical structures, and their proximity to other biological agents such as tissue or tumor boundaries, or blood vessels~\cite{Greenwald2024, Bich2019}. This surrounding microenvironment, or \emph{niche}, affects a cell's gene expression profile and therefore its identity and fate~\cite{Adema2024, Asp2020, Jewell2023}. This holds across scales: from location-dependent macrophage phenotypes in the tumor microenvironment~\cite{Huang2019} to laminar organization driving cortical function~\cite{Qian2025}.
In each case, molecular measurements without their spatial frame yield an incomplete, sometimes misleading, picture of biological reality~\cite{Bressan2023, Zormpas2023}.



\ST closes this gap by retaining both molecular and spatial information~\cite{Tian2022, Chen2023}. However, this creates a new visualization problem: analysts must now reason jointly across molecular measurements, spatial arrangements, tissue morphology, segmentation boundaries, and derived annotations.

\subsection{\ST Technologies and Data Structures}

\begin{figure*}[t]
 \centering
  \includegraphics[width=0.80\textwidth]{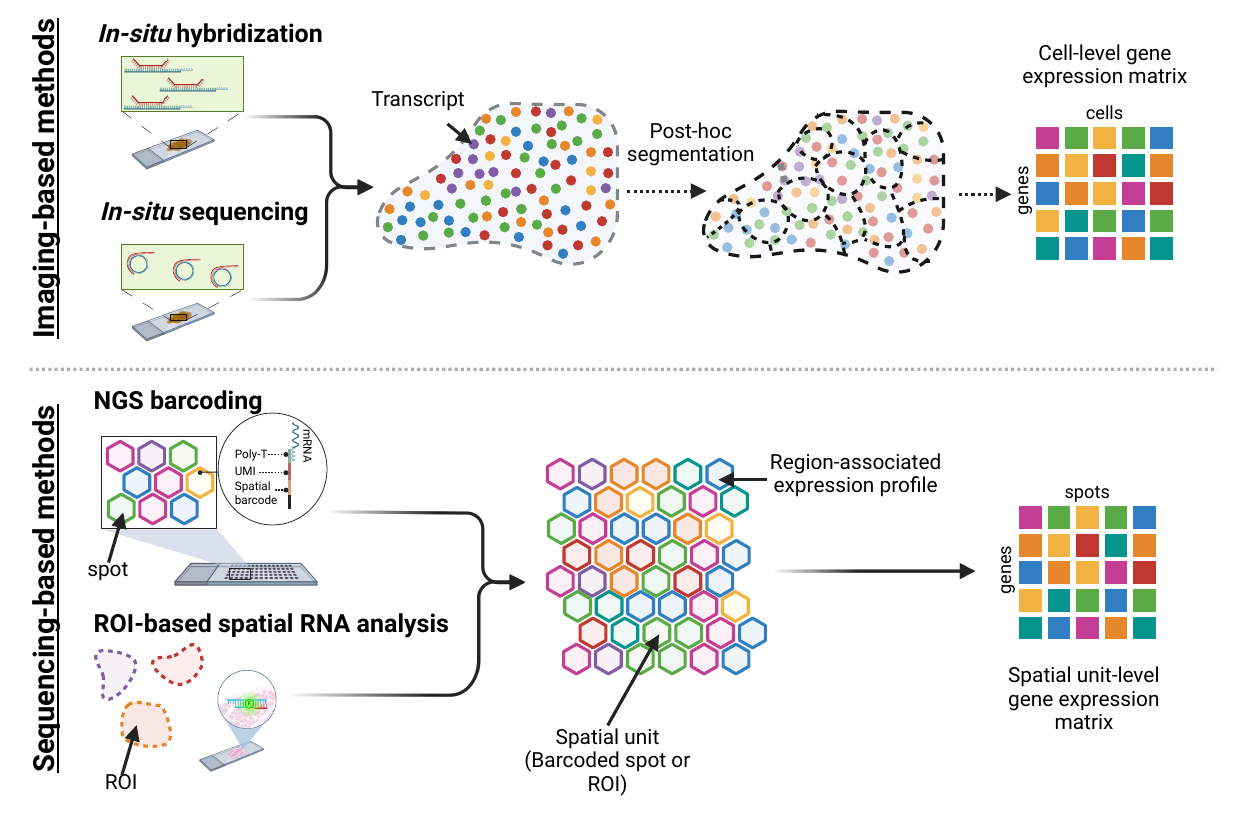}
  \caption{\ST technology families and their data structures. \emph{Imaging-based} methods (ISH, ISS) directly localize individual transcripts, producing point clouds without inherent cell identities. \emph{Sequencing-based} methods (barcoded arrays, ROI-based) associate measurements with predefined spatial units, yielding expression matrices directly. Adapted from~\cite{Jain2024, Rao2021, Lim2025}; Figure created with BioRender.}
  \label{fig:spatial_transcriptomics}
\end{figure*}

\ST technologies fall into two major families based on their detection strategy: \emph{imaging-based} and \emph{sequencing-based}, each presenting a distinct set of representational trade-offs (Figure~\ref{fig:spatial_transcriptomics})~\cite{Jain2024, Zormpas2023}. From a data perspective, the choice of technology determines the structure, resolution, and modality of the data that visualization tools must contend with.

Imaging-based methods directly visualize RNA molecules using fluorescence microscopy, offering sub-cellular resolution but requiring a pre-selected gene panel~\cite{Yue2023}. \emph{In situ hybridization} (ISH) methods use combinatorial barcoding across imaging rounds to detect hundreds to thousands of transcripts at single-molecule sensitivity~\cite{Chen2015, Fortner2024MultiplexedST, Cilento2024}, while \emph{in situ sequencing} (ISS) methods read short barcoded sequences directly within the tissue~\cite{Ke2013, Wang2018}. Both approaches ultimately produce a spatial point cloud of detected transcripts rather than measurements inherently associated with individual cells. Therefore, \emph{cell segmentation} becomes a critical analytical step. This process of inferring cell boundaries is typically based on nuclear staining and watershed-style propagation~\cite{Marks2025}.


Sequencing-based approaches, instead, capture RNA from spatially defined locations and identify transcripts by next-generation sequencing, enabling whole-transcriptome profiling without pre-selecting gene targets~\cite{Shi2024}. \emph{Array-based} platforms place tissue on a grid of barcoded capture spots, yielding an expression matrix anchored to a structured grid. The resolution of this grid has improved over successive platform generations, from several cells per spot down to sub-cellular spot sizes~\cite{Sthl2016, Oliveira2025}. In contrast, \emph{region-of-interest} (ROI) methods let a researcher select and sequence specific tissue regions independently, yielding whole-transcriptome profiles tied to coarse, analyst-defined polygons~\cite{Jain2024, Park2023}.

Despite their differences, all \ST workflows produce a common tripartite data organization. The \emph{topological layer} provides spatial coordinates that anchor every observation to a physical tissue location, with a topology that varies by technology: structured grids of capture spots (array-based methods), irregular point clouds of individual transcript molecules (imaging-based methods), or analyst-defined polygons (ROI methods). The \emph{transcriptional layer} takes one of two forms depending on technology: for sequencing-based methods, it is a matrix $\mathbf{X} \in \mathbb{N}_{0}^{n \times m}$ of raw integer counts produced directly during acquisition; for imaging-based methods, it is a matrix $\mathbf{X} \in \mathbb{R}_{\geq 0}^{n \times m}$ derived only after cell segmentation aggregates transcript detections from $\mathcal{P}$ into per-cell counts. In both cases $n$ denotes the number of spatial units and $m$ the number of genes; normalized downstream representations take values in $\mathbb{R}^{n \times m}$. The \emph{morphological layer} is a high-resolution raster tissue image, typically a brightfield hematoxylin and eosin (H\&E) stain or immunofluorescence image, providing the histological context indispensable for biological interpretation~\cite{Zormpas2023}. Alignment (\emph{registration}) between the coordinate and image layers is a prerequisite for valid inference; misalignment propagates downstream to all spatial analyses~\cite{Zhang2025FromST}. Beyond these three core layers, downstream analyses produce additional derived structures: segmentation masks, spatial neighborhood graphs, cell-type annotation vectors, and cell--cell communication networks. Datasets frequently also carry experimental metadata across conditions and donors. We discuss each of these as distinct visualization targets in subsequent sections. 


\subsection{Data Characteristics and Their Visualization Implications}
\label{sec:data-characteristics}

Each property of \ST data introduced above creates its own constraint on visualization design. While techniques exist for high dimensionality, spatial embedding, or multimodality individually, \ST data combines all of them simultaneously: it is high-dimensional, spatially embedded, multimodal, uncertain, and large-scale. A single experiment may profile tens of thousands of genes across hundreds of thousands of spatial units, involving gigapixel tissue images and millions of transcript points that must all be jointly registered. The trade-off between finding molecular patterns and staying spatially faithful is compounded by the fact that biology itself is multiscale: meaningful phenomena occur at the level of molecules, cells, tissue regions, and whole sections all at once, and a pattern that is obvious at one scale can be invisible, or misleading, at another. 

Further, expression is not independently distributed across space: nearby units tend to exhibit correlated profiles reflecting shared microenvironmental signals, so visualizations that treat spatial units as exchangeable discard the very dependency structure that is often the primary object of analysis.

Uncertainty is present at every level. In sequencing-based methods, a zero in the expression matrix may reflect biological inactivity or simply result from low capture efficiency~\cite{Wang2022,Song2023}; in imaging-based methods, segmentation leaves roughly 30\% to 40\% of detected molecules unassigned~\cite{Salas2025}. Cell-type annotations, trajectory inferences, and spatial statistics carry further uncertainty that is rarely propagated to visualization. 

Additionally, datasets can exceed hundreds of thousands of spatially indexed cells with gigapixel images, requiring level-of-detail strategies for interactive use. When time is present, through developmental atlases, perturbation experiments, or RNA velocity inferences~\cite{Swift2021}, it introduces ordered comparisons that further increase the complexity of visual encodings.

These coupled properties frame \ST as a demanding visualization design domain. Conventional single-cell tools such as scatter plots of dimensionally reduced data, heatmaps, or trajectory diagrams provide partial views but cannot jointly address spatial structure, multimodality, uncertainty, and scale. The coding framework and survey that follow are organized around this challenge.

\section{Methods}

\begin{figure*}[!t]
\centering
  \includegraphics[width=0.85\textwidth]{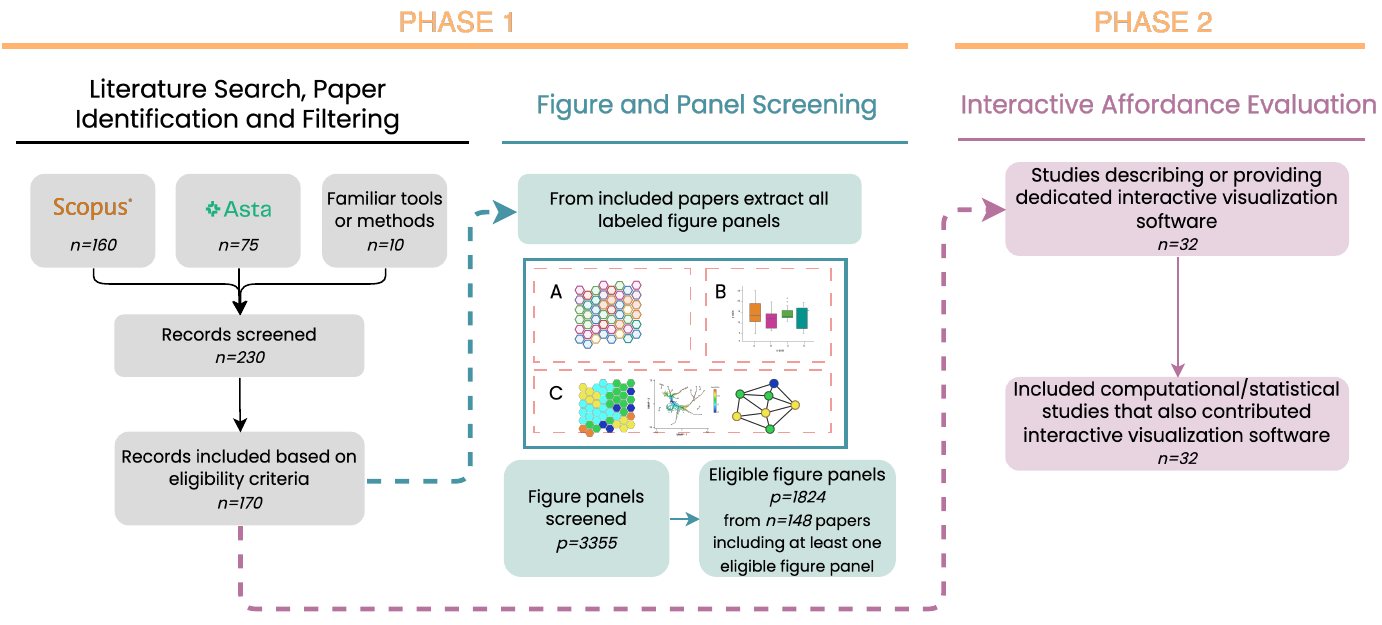}
  \caption{Literature search, paper identification and filtering. Phase~1 focused on paper identification, filtering, and panel extraction. Initial records were retrieved from Scopus ($n=160$) and the Allen Institute's Asta semantic search platform ($n=75$). After removing 15 duplicates, 220 unique records were screened for relevance against our inclusion criteria, excluding 59 records as out of scope. Of the remaining 161, one could not be retrieved, yielding a primary stream of 160 eligible reports. This was supplemented with 10 additional studies that had been identified independently by the authors but were not retrieved by the search query, yielding a final corpus of 170 included studies. From these papers, all labeled figure panels were extracted and classified based on explicit inclusion and exclusion criteria, resulting in $1{,}824$ qualifying and $1{,}531$ non-qualifying panels. Phase~2 comprised the interactive affordance evaluation, wherein the 32 included studies that described or contributed dedicated interactive visualization software were identified. The interactive capabilities of these tools were coded and analyzed using the taxonomy of Yi \etal~\cite{Yi2007}}.
  \label{fig:methods}
\end{figure*}

This study investigates visualization practices in ST, in particular how the field visually encodes, reasons about, and communicates spatially resolved biological data. The analysis is based on two methodological components: the construction of a literature corpus and our coding procedure (\Cref{fig:methods}). 

\subsection{Scope and Corpus Construction}

New computational methods in \ST frequently produce spatially anchored outputs, such as spatially variable gene scores or niche assignments, for which an established visual convention does not necessarily exist. Therefore, introducing a new method may implicitly introduce a new visual encoding as well. To capture these emerging design choices, we explicitly review primary methodological literature. This decision addresses a coverage limitation, as dedicated visualization software integrates only a subset of the field's computational methods. Examining primary methodology papers allows us to analyze unique, approach-specific visual designs at their point of origin, including specialized representations that have not yet been, and may never be, incorporated into visualization software. Within these papers we treat the static publication figure as the primary unit of analysis. Static figures reflect how researchers visualize their data to reach and communicate key findings, and it is in this process that encoding decisions become concrete, as a researcher commits to a color scale, a spatial layout, or a level of abstraction. Furthermore, static figures offer a stable and accessible record for systematic analysis, even though they necessarily represent a fixed snapshot of a broader analytical process. By examining this visual record, we characterize the strategies used to communicate spatial findings and evaluate how visual reasoning is performed across the field.

We therefore conducted a systematic review of static figures published in papers describing computational and statistical methods for spatially resolved transcriptomics data analysis. Some of the methods papers that matched our search happened to also include, contribute, or discuss dedicated, often interactive, visualization software; where this was the case, we also considered that dedicated software in our analysis. 

We assembled a corpus of peer-reviewed articles and preprints published between 2018 and 2025. The lower bound reflects the advent of high-throughput \ST technologies that triggered the current wave of methodological development. We included two broad categories of work: (1)~computational and statistical methods that introduce novel spatial quantities or representations, and (2)~tools designed explicitly for static or interactive exploration of spatially resolved biological data. Our search combined keyword-based queries in Scopus with AI-assisted semantic search via the Allen Institute's Asta platform (\url{https://asta.allen.ai}), which helped overcome the limitations of keyword-based retrieval. We provide the complete set of search queries used in the Supplemental Material.

We treated paper and panel selection as distinct stages of corpus construction, each governed by explicit criteria that reflect the analytical scope of the study.
At the paper level, we included papers whose figures visually encode spatial coordinates, tissue context, or spatial biological structure. These papers included both those whose primary contribution was either analytical (e.g., introducing a new spatial statistical test) or presentational (e.g., contributing a tool for visual exploration). We excluded papers when visualization was exclusively used for the purpose of benchmarking---that is, when figures were only used to support quantitative performance comparisons across methods rather than to communicate biological interpretation. We also excluded wet-lab protocol papers, book chapters, and papers with purely temporal or non-spatial visualizations.


At the panel level, we included figure panels when they visually represented spatially resolved biological information. This inclusion criterion could be met through either explicit spatial coordinates or tissue context, or through analytical abstractions derived from spatial organization, such as spatial domains, cellular neighborhoods, proximity relationships, or other spatially defined structures. We excluded panels when they did not contribute to the visual communication of spatial biological information, including method-evaluation benchmarking plots, schematic workflow diagrams, and pipeline illustrations. Rather than omitting excluded panels during annotation, we assigned them an explicit exclusion code, enabling inter-rater assessment and potential analysis in future studies while excluding them from this study. Of the 170 papers admitted in Phase~1 (Fig. \ref{fig:methods}), 22 contributed exclusively non-qualifying panels under these criteria.  These 22 papers remain part of the corpus, but do not contribute panel-level data to the results reported here. After screening, 148 papers and 1,824 panels qualified for inclusion and form the basis of the analyses presented in Section~\ref{sec:results}.

\subsection{Coding Process}
\label{subsec:codding_process}

We annotated each qualifying figure panel against a structured coding framework organized around three coupled axes: \taxaxis{what} data are visually encoded, \taxaxis{why} the visualization is used (the biological analytical intent it supports), and \taxaxis{how} the data are visually represented. This framework is grounded in Munzner's nested model~\cite{munzner2015visualization} and operationalized as three interdependent taxonomic components---a \taxaxis{data} axis, a \taxaxis{task} axis, and a \taxaxis{visualization} axis---described in Section~\ref{sec:taxonomy}. We developed and refined the coding framework iteratively: we applied initial category definitions to a subset of the corpus, resolved ambiguous cases through author discussion, and then applied the revised framework to the full corpus.


When a single panel combined multiple visual representations, we applied codes to each representation independently.
For instance, for a heatmap presented alongside a spatial gene-expression map within the same labeled subfigure, we coded the heatmap and spatial visualizations independently. Additionally, we allowed taxonomy dimensions to differ in cardinality constraints. We treated modality, resolution, experimental condition, spatial layout, abstraction level, and comparative design as primary dimensions where a dominant category was expected and a single label assigned. By contrast, we designed data components, visualized elements, scalability strategies, and contextualization elements to accept multiple labels, reflecting the frequent co-occurrence of several data types, mark types, or communicative devices within a single panel. 

Two annotators independently coded a random subset of 11 papers from the corpus (comprising 259 figure panels) during an initial calibration phase. After these papers were coded by both annotators, the authors reconciled disagreements to refine the framework, which a primary coder then applied to the full corpus, with targeted refinements as unanticipated cases arose.

Beyond the structured coding framework labels, annotating biological intent revealed that the same high-level task category was able to serve qualitatively different scientific purposes depending on context. To preserve these distinctions, we additionally recorded, for each panel, a concise natural-language abstraction of the research question the figure panel was designed to answer. We recorded these  abstractions at the time of annotation. To synthesize these raw annotations into interpretable structure, we grouped the research question abstractions into higher-level thematic categories using an LLM-assisted synthesis process. We used \texttt{GPT-5.4-05-2026} as the underlying model. The model served as a clustering and summarization aid: it proposed candidate groupings of semantically related research questions, which the authors then manually reviewed, corrected, merged, or split. The LLM then applied the finalized category definitions to classify each research question abstraction, using a structured human-in-the-loop procedure: it produced classifications for different research questions in review batches, an author checked the classifications and provided corrections to update the coding rules, and we propagated the revised rules to subsequent batches before performing a full classification pass.

Although this study is primarily grounded in the analysis of static publication figures, we also examined the interactive tools identified within our corpus of included papers. Because interactivity is often described in text at the level of the software system rather than being observable in any single static figure panel, we conducted this analysis at the paper or tool level rather than the panel level. During the paper-selection stage, we identified those papers that explicitly described interactive features. For each such paper or tool, we coded the reported interaction capabilities using the taxonomy proposed by Yi \etal~\cite{Yi2007}, which distinguishes seven categories of interaction intent: \emph{abstract/elaborate} (adjusting the level of detail), \emph{reconfigure} (changing the spatial arrangement of visual elements), \emph{encode} (altering the visual mapping), \emph{select} (marking items of interest), \emph{filter} (suppressing unselected items), \emph{explore} (panning or navigating to reveal additional data), and \emph{connect} (highlighting relationships across views). We characterized the types of interaction support currently emphasized in the tools and related these interaction types to our \taxaxis{task} axis (i.e., the \taxaxis{task} axis developed through panel-level coding).

\section{A What-Why-How Framework for \ST Visualization}
\label{sec:taxonomy}

\begin{figure*}[p]
  \includegraphics[width=\textwidth]{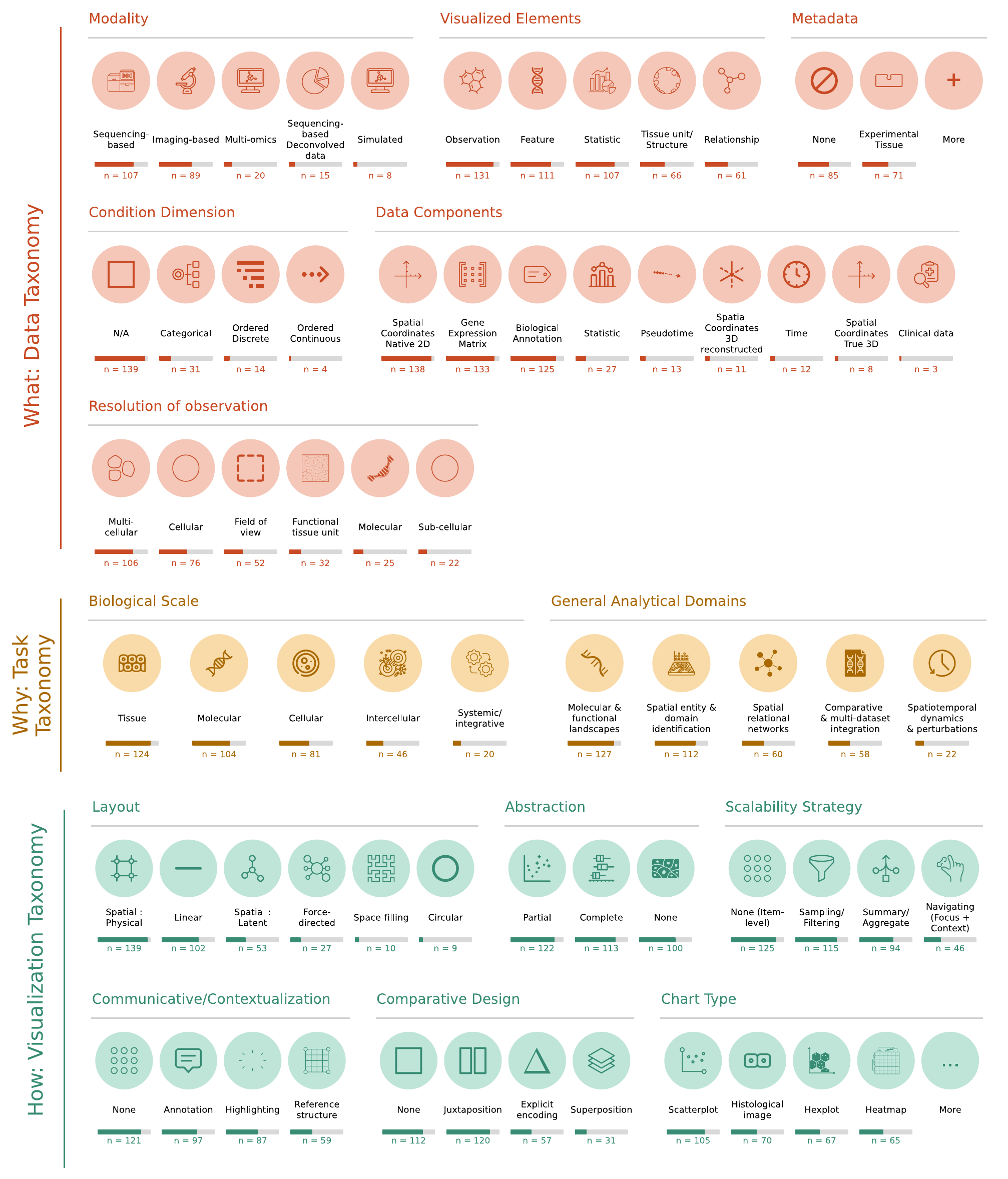}
  \caption{\taxaxistext{Data}, \taxaxis{task}, and \taxaxis{visualization} axes of the coding framework used for \ST figures. The \taxaxis{data} axis characterizes what is visually encoded through six dimensions: \taxdim{modality}, \taxdim{resolution}, \taxdim{condition}, \taxdim{data-components}, \taxdim{metadata} and \taxdim{visualized-elements}. The \taxaxis{task} axis captures why a visualization is used through the \taxdim{biological-scale} and \taxdim{analytical-domain} dimensions. The \taxaxis{visualization} axis describes how \ST data are visually expressed through \taxdim{layout}, \taxdim{abstraction}, \taxdim{comparison}, \taxdim{scalability}, \taxdim{contextualization}, and \taxdim{chart-type}. All $n$ values denote paper-level counts (number of papers with $\geq 1$ qualifying panel exhibiting the code).}
    \label{fig:taxo}
\end{figure*}

Characterizing visualization practice in a rapidly evolving scientific domain requires a conceptual framework that goes beyond cataloging chart types or software tools.
Therefore, we ground our coding framework in Munzner's nested model and its \emph{What--Why--How} framework~\cite{munzner2015visualization}.
We operationalize \taxaxis{what}, \taxaxis{why}, and \taxaxis{how} as three coupled taxonomy components. The \taxaxis{data} axis describes \emph{what} is being represented, the \taxaxis{task} axis captures the analytical \emph{intent} motivating the visualization, and the \taxaxis{visualization} axis captures the \emph{visual design choices} through which data and intent are expressed (\Cref{fig:taxo}).

As with the results reported in \Cref{sec:results}, counts $(n)$ in \Cref{fig:taxo} reflect paper-level prevalence rather than raw panel-level occurrence: a paper is counted once toward a given code if at least one qualifying panel exhibits that code, regardless of how many panels within that paper repeat it. This prevents a single paper with many similar panels from inflating a category relative to a paper contributing only one panel. Because labels are still assigned per panel, a paper whose panels span multiple categories within the same dimension (e.g., both imaging- and sequencing-based figures) contributes to each; category counts therefore need not sum to 148 papers or 1,824 panels.

\subsection{\taxaxis{what}: Data Components in \ST}

The \taxaxis{data} axis characterizes the information encoded in each figure panel. The \taxdim{modality} dimension distinguishes \taxcode{imaging}, \taxcode{sequencing}, and \taxcode{mapped} data. The \taxdim{resolution} dimension captures the biological granularity of the information, from \taxcode{molecular} and \taxcode{subcellular} through \taxcode{cellular}, \taxcode{multicellular}, and \taxcode{tissue-unit} levels up to the \taxcode{full-fov}. The \taxdim{condition} dimension records whether the data involve a comparative structure, comprising \taxcode{categorical}, an \taxcode{ordered} (\taxcode{discrete} or \taxcode{continuous}), or \taxcode{neither}.

The axis further records which \taxdim{data-components} are visually present: \taxcode{expression}, \taxcode{coordinates} (\taxcode{native-2d}, \taxcode{true-3d}, or \taxcode{reconstructed-3d}), \taxcode{rna-velocity}, \taxcode{temporal}, and \taxcode{annotations}. It also distinguishes which \taxdim{visualized-elements} appear: \taxcode{observations} (cells), \taxcode{features} (genes or pathways), \taxcode{relationships}, or \taxcode{aggregates}. The axis also records \taxdim{metadata} presence when experimental conditions or donor information are visually indicated. Together, these dimensions establish the representational vocabulary on which the \taxaxis{task} and \taxaxis{visualization} axes build.

\subsection{\taxaxis{why}: Biological Task Structure}
\label{subsection:taxonomy-why}

The \taxaxis{task} axis characterizes the biological reasoning that a visualization is designed to support. We characterize the biological question that each figure panel answers. We do so through two complementary dimensions: first, by recording the \taxdim{biological-scale} at which the visual reasoning operates and, second, by abstracting the specific research question addressed by the panel into a \taxdim{analytical-domain}.

Within the \taxdim{biological-scale} dimension, we distinguish five codes. \taxcode{molecular} tasks concern gene regulation, pathway activity, co-expression structure, and other molecular programs. \taxcode{cellular} tasks address cell identity, marker gene expression, cell states, and state transitions. \taxcode{intercellular} tasks focus on relationships among cells, including cell--cell communication, spatial neighborhoods, niches, and proximity-based interactions. \taxcode{tissue} tasks concern spatial domains, anatomical organization, tissue layers, and other structures that emerge at scales above individual cells. Finally, \taxcode{systemic} tasks span multiple conditions, modalities, samples, or biological scales to synthesize heterogeneous biological evidence.

This scale-based coding captures the level of biological organization at which a visualization operates, but scale alone does not fully describe a figure's analytical intent. For example, a spatial tissue map colored by cell type and a spatial tissue map colored by pathway activity may use a similar visual form, yet support different biological questions: one concerns overall cell identity and organization, whereas the other concerns the spatial pattern of a particular molecular function. 

To preserve these distinctions, we also recorded a concise natural-language abstraction of the research question addressed by each subfigure, which we later clustered into more abstract groups and translated into the five \taxdim{analytical-domain} codes shown in the right-hand panel of \Cref{fig:taxo}: \taxcode{task-molecular}, \taxcode{task-spatial}, \taxcode{task-relational}, \taxcode{task-spatiotemporal}, and \taxcode{task-comparative}. This clustering process is described in more detail in \Cref{subsec:codding_process}.

\subsection{\taxaxis{how}: Visualization Design Strategies}

The \taxaxis{visualization} axis characterizes representational strategy across six dimensions. Extending the taxonomy proposed by Nusrat \etal~\cite{Nusrat2019}, \taxdim{layout} encodes how visual elements are spatially arranged, distinguishing \taxcode{physical} layouts from \taxcode{latent} projections, as well as \taxcode{linear}, \taxcode{circular}, \taxcode{space-filling} and \taxcode{force-directed} arrangements. The \taxdim{abstraction} dimension captures the degree to which physical reality has been simplified, from \taxcode{realistic} representations that preserve tissue morphology and photographic detail, through \taxcode{partial} representations that maintain spatial relationships but reduce visual complexity, to representations classified as \taxcode{abstract} that retain no spatial reference.

The \taxdim{comparison} dimension records how visual comparisons are implemented, following the framework established by Gleicher \etal~\cite{Gleicher2011}: \taxcode{juxtaposition} (separate panels or facets), \taxcode{superposition} (layered encodings on a shared coordinate space), \taxcode{explicit} (difference or ratio encoded directly as a visual variable), or \taxcode{no-comparison}. In addition, the \taxdim{contextualization}  dimension records communicative additions that support interpretation without changing the primary encoding, including \taxcode{text-graphics}, \taxcode{highlighting} through contrast or desaturation, and \taxcode{references} such as tissue boundaries, anatomical landmarks, and scale bars. Loosely based on considerations by Gleicher \etal~\cite{Gleicher2018}, we included a \taxdim{scalability} dimension, capturing how data size and complexity are managed through \taxcode{item-level}, \taxcode{aggregation} into bins or regional summaries, \taxcode{sampling}, or \taxcode{focus-context}. Finally, \taxdim{chart-type} captures the specific visualization form used (e.g., \taxcode{scatterplot}, \taxcode{heatmap}, or \taxcode{violin}).

\begin{figure}[hbp]
  \includegraphics[width=\columnwidth]{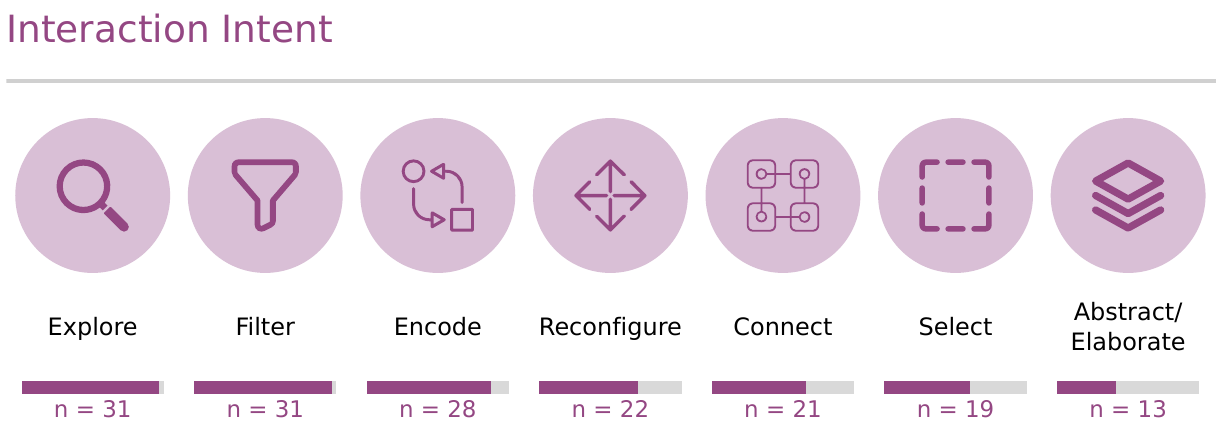}
  \caption{Frequency of codes in the \taxdim{interaction-intent} dimension across the 32 surveyed \ST visualization tools, coded using the taxonomy of Yi \etal~\cite{Yi2007}.}
  \label{fig:interactivity_taxonomy}

\end{figure}

Beyond these six static-figure dimensions, we separately characterize interactivity for the subset of surveyed works that contribute dedicated visualization software. We annotate each such tool at the system level using the taxonomy of interaction intent proposed by Yi \etal~\cite{Yi2007}, which distinguishes seven codes for the \taxdim{interaction-intent} dimension: \taxcode{select}, \taxcode{explore}, \taxcode{reconfigure}, \taxcode{encode}, \taxcode{abstract-elaborate}, \taxcode{filter}, and \taxcode{connect} (\Cref{fig:interactivity_taxonomy}). This lets us relate the representational strategies captured by the dimensions described above to the interactive capabilities that accompany them where such tooling exists and ask whether the interaction affordances a system offers align with the analytical intent of the task it supports.

\section{Results}
\label{sec:results}

Across 148 papers, we identified 1,824 qualifying figure panels and coded them along the \taxaxis{data}, \taxaxis{task}, and \taxaxis{visualization} axes; for papers that reported interactive capabilities for data visualization, we additionally coded supported interaction options by tool. The coded corpus can be explored through our web-based survey browser (\url{https://denisseram.github.io/STAR-spatialsc/}).

We organize this section around the five \taxdim{analytical-domain} codes introduced in \Cref{subsection:taxonomy-why}, each of which anchors one subsection below: \Cref{sec:molecular} (\taxcode{task-molecular}), \Cref{sec:spatial entity} (\taxcode{task-spatial}), \Cref{sec:relational} (\taxcode{task-relational}), \Cref{sec:spatiotemporal} (\taxcode{task-spatiotemporal}), and \Cref{sec:comparative} (\taxcode{task-comparative}). Within each subsection, we break down the broad domain into its specific tasks, responding to a cluster of research questions identified during coding. 

All percentages below are \emph{paper-level prevalence}: the share of papers associated with a given research question that have at least one qualifying panel exhibiting a category. This prevents papers containing many panels of the same visualization type from disproportionately influencing the observed patterns. For example, a paper containing multiple similar scatterplots contributes once to the presence of that visualization type rather than once per panel. Categories are not mutually exclusive, so figures within a task need not sum to 100\%, and a paper can register simultaneously as, for example, using both a \taxcodeformat{physical spatial} layout and a \taxcode{abstract} representation in different panels.

\subsection{Molecular \& Functional Landscapes: Visualizing Continuous Expression and Pathways}
\label{sec:molecular}

\begin{figure*}[t]
  \centering
  \includegraphics[width=\textwidth]{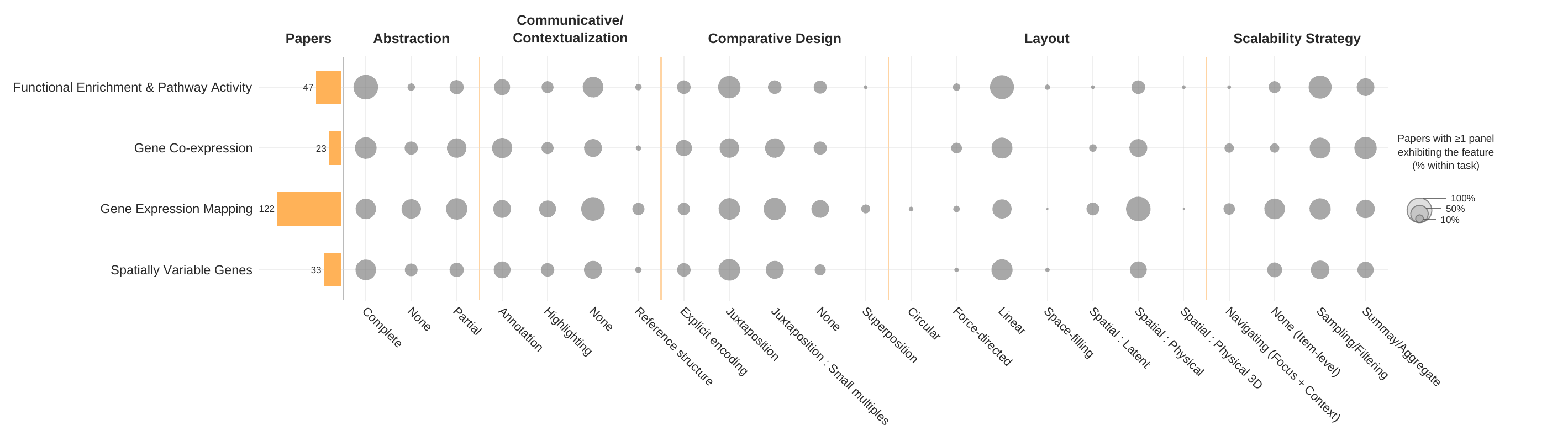}
  \caption{\label{fig:molecular_functional_landscapes} Visualization design patterns associated with \taxcode{task-molecular} tasks.}
\end{figure*}

The molecular scale is where \ST begins: before any claim about cell types, tissue domains, or signaling relationships can be made, analysts must determine: (\textit{i})~how individual genes are distributed across tissue space (\taxtask{gene-expression}),
(\textit{ii})~which features vary systematically with position (\taxtask{svg-discovery}), (\textit{iii})~how genes co-vary (\taxtask{coexpression}), and (\textit{iv})~what functional programs those patterns imply (\taxtask{pathway-enrichment}).
The visualization idioms of choice shift along this chain of sub-tasks, moving from spatially explicit displays of raw expression toward the more abstract, list- and network-based representations that dominate pathway-level reasoning (see counts in \Cref{fig:molecular_functional_landscapes}).

    \subsubsection{Gene Expression Mapping}
    \label{sec:targeted-expression}

    \taxtaskformat{Gene expression mapping} is the most frequent task in our corpus ($n=122$ papers, 653 panels), and is used to check whether specific molecular signals form spatially coherent structures. Its data substrate is dominated by \taxcodeformat{gene expression matrices} (87.7\%, $n=107/122$) anchored to \taxcode{native-2d} spatial coordinates (92.6\%, $n=113/122$), with \taxcode{annotations} co-occurring in 83.6\% of papers ($n=102/122$).

\begin{figure}[tbp]
      \centering
      \includegraphics[width=0.38\textwidth]{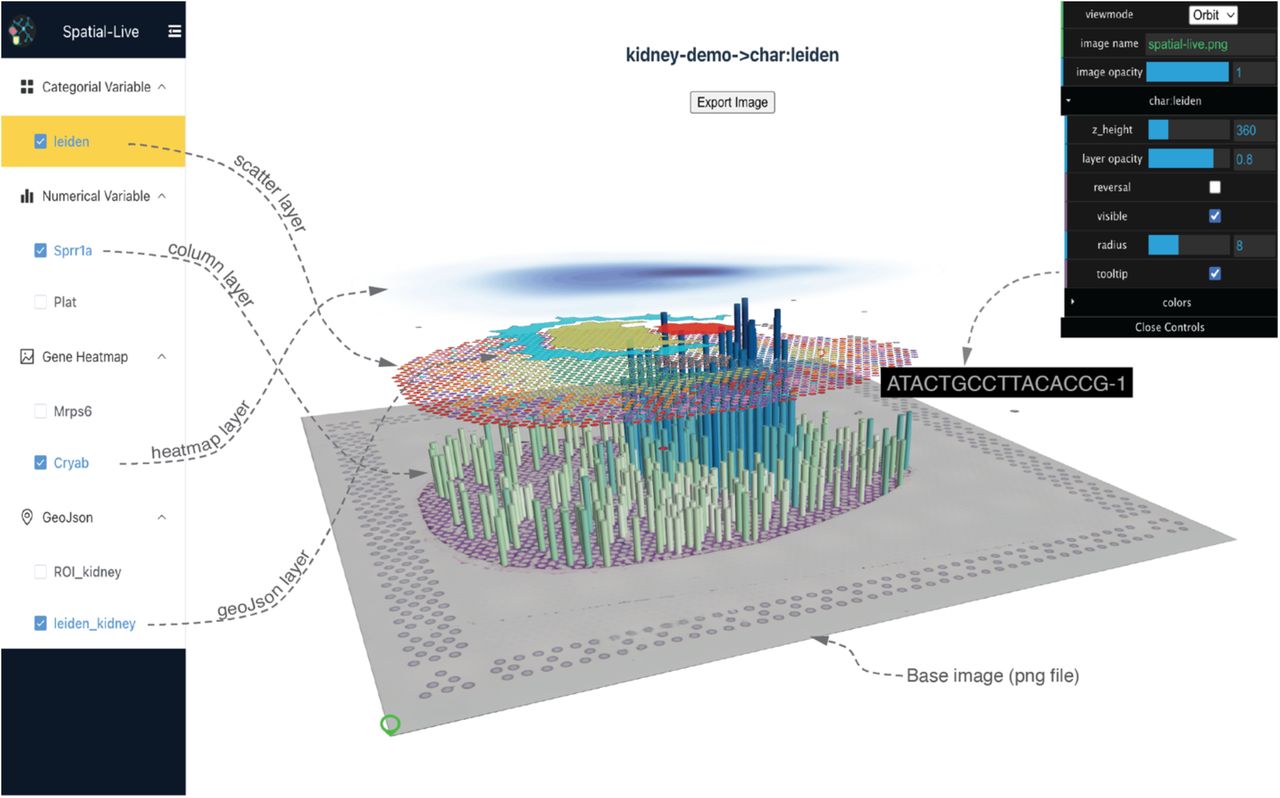}
      \caption{%
          Example of an interactive visual environment mapping numerical expression magnitudes to bar height (\textit{column layer}), superimposed over cell clusters (\textit{scatter layer}) and a registered histology image. Reproduced from Ye \etal~\cite{Ye2023}. Panels adapted with permission from the authors.}
    \label{fig:spatial_3d_columns}
    \end{figure}

    The dominant visual archetype is the \taxcodeformat{physical spatial} layout, used in 89.3\% of papers ($n=109/122$), most often as \taxcodeplural{scatterplot} or \taxcodeplural{hexplot} mapping continuous expression to sequential color. Color, however, has a ceiling: when several variables need to be shown at once, it quickly becomes perceptually ambiguous. \Cref{fig:spatial_3d_columns} shows one way the field works around this limit, extruding 3D bar glyphs from the tissue plane so that magnitude is encoded as bar height while color is freed up to represent cell cluster identity instead. \taxcodeformat{linear, non-spatial} companion summaries (violin or \taxcodeplural{bar-plot}) offer a simpler workaround and are common in their own right, present in 54.9\% of papers ($n=67/122$).

\begin{figure}[b]
      \includegraphics[width=0.45\textwidth]{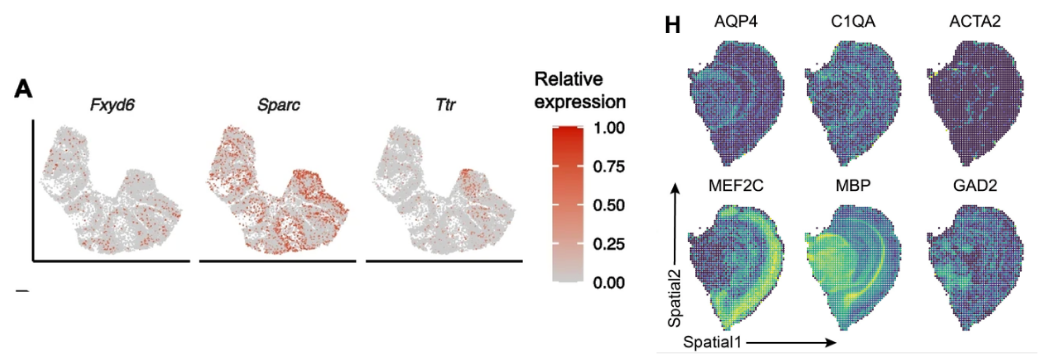}
      \caption{
      Representative multi-panel comparative layouts showing continuous expression across different genes within the same tissue substrate. (Left) Spatial \taxcodeplural{scatterplot} preserving physical positioning (Imaging-based, reproduced from Yu \& Li~\cite{yu2024}). (Right) Gridded hexagonal spot matrices showing discrete regional binning (Visium, reproduced from Ruan \etal~\cite{ruan2024}). Panels adapted under CC BY 4.0.}
      \label{fig:gene_scatter_hexplot}
    \end{figure}

    A second pressure point is comparison across gene expression patterns. The field relies heavily on small-multiple \taxcode{juxtaposition} (used in 73.8\% of the papers in the subset, $n=90/122$), placing single-gene panels side by side rather than folding multiple genes into one view. \Cref{fig:gene_scatter_hexplot} illustrates this for both \taxcode{imaging} and Visium-style spot data. However, which genes make it into that grid is rarely decided by the data itself: \taxcode{sampling} is applied in 67.2\% of papers ($n=82/122$), typically through a manually curated marker-gene set. This manual selection, together with the color-blending limits of dense small multiples, remains an open challenge.

    \subsubsection{Spatially Variable Genes}
    \label{sec:svg}


    Detection of \taxtaskformat{spatially variable genes} (SVGs) transforms a high-dimensional expression matrix into a ranked list of spatially informative genes, a process tightly connected to the targeted-expression analyses above. Unlike  \taxtask{gene-expression}, which asks \emph{where} a known gene is expressed, SVG identification asks \emph{which} genes are worth mapping at all.

    With $n=33$ papers (66 panels), this sub-task is substantially less common than targeted expression. The shift from observation to ranking that it demands is reflected in the data the figures encode: \taxcodeformat{gene expression matrices} appear in 90.9\% of papers ($n=30/33$) and \taxcode{native-2d} coordinates in 72.7\% ($n=24/33$). \taxcode{sequencing} methods dominate (72.7\%, $n=24/33$), consistent with whole-transcriptome SVG detection being built primarily for platforms capable of measuring a larger gene set; \taxcode{imaging} methods, constrained to pre-selected gene panels, appear in only 30.3\% ($n=10/33$).

\begin{figure}[htbp]
       \centering
       \includegraphics[width=0.65\columnwidth]{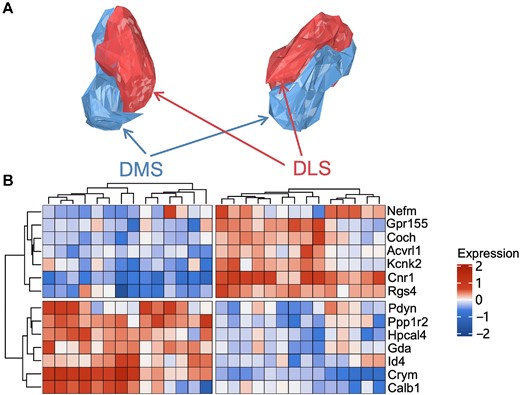}
       \caption{Example of anatomical tissue contextualization paired with a linear feature map. (A)~Spatial tissue reference showing localized regional structures. (B)~Companion heatmap displaying continuous gene expression levels across the corresponding regions. Reproduced from Cao \etal~\cite{cao2023}. Panels adapted under CC BY 4.0.}
       \label{fig:svg_reference_heatmap}
     \end{figure}

    This evaluative framing shapes figure design. \taxcodeformat{linear, non-spatial} layouts outnumber \taxcodeformat{physical spatial} layouts here (66.7\% vs.\ 42.4\%, $n=22/33$ vs.\ $n=14/33$), as domain experts frequently display ranked gene lists or statistical scores in linear format, reserving spatial maps for illustrating exemplary top hits, as shown in \Cref{fig:svg_reference_heatmap}. Indeed, 63.6\% of papers ($n=21/33$) use \taxcodeformat{complete abstraction}, which is consistent with the task's emphasis on ranking over grounding.

    \subsubsection{Gene Co-expression}
    \label{sec:co-expression}

    Once SVGs are identified, a natural next question is whether they co-vary systematically. Genes may be part of coherent spatial programs, that is, modules of genes whose expression rises and falls together across tissue space. Co-expression and co-localization analysis reduces the expression matrix to a smaller set of these latent spatial programs. Because a module is defined jointly by its membership (which genes belong to it) and its behavior (where in the tissue it is active), visualizing such a module is often a two-part task, pairing a gene-level view with a tissue-level view.

    This is the least common molecular sub-task in the corpus ($n=23$ papers, 53 panels), and its data substrate reflects its dual nature: \taxcodeformat{gene expression matrices} appear in 95.7\% of papers ($n=22/23$), while \taxcode{native-2d} coordinates appear in 60.9\% ($n=14/23$). The difference indicates that module analysis is also frequently performed in a representational space once removed from the tissue itself. \taxcode{annotations} still co-occur in 69.6\% of papers ($n=16/23$), used to check whether a discovered module maps onto a known anatomical structure or cell type, a way of tethering the abstraction back to biology.
    
\begin{figure}[btp]
      \centering
      \includegraphics[width=0.85\columnwidth]{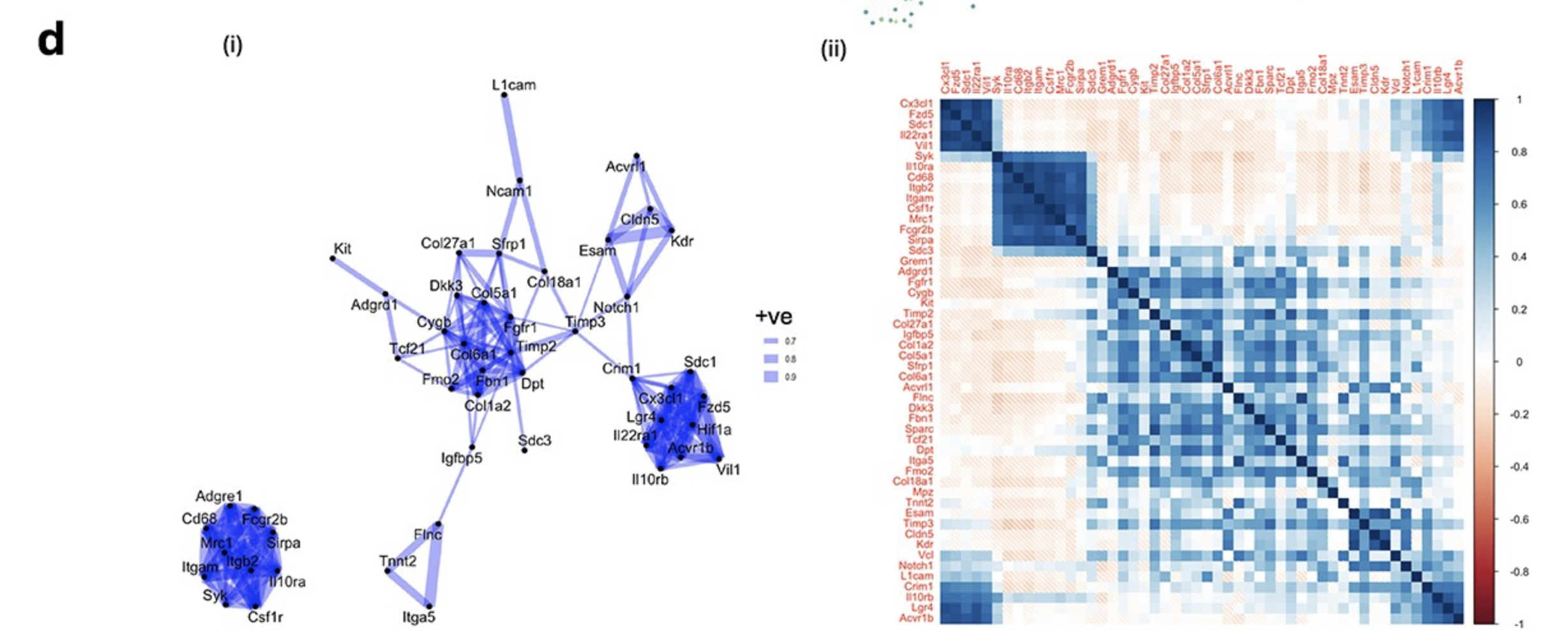}
      \caption{Example of a co-expression layout juxtaposing an estimated gene co-regulation network graph with its corresponding linear correlation matrix. Reproduced from Egbon \etal~\cite{Egbon2025}. This work is a U.S. Government work and is in the public domain in the United States.}
    \label{fig:coexpression_modules}
    \end{figure}

    This two-part logic directly shapes figure design. \taxcodeplural{heatmap} are the dominant chart type (56.5\%, $n=13/23$), followed by \taxcodeplural{scatterplot} (43.5\%, $n=10/23$), pairing a gene-membership view (\emph{which genes}), with a spatial map of module activity (\emph{where}). These two views are typically shown as \taxcode{juxtaposed} panels (56.5\%, $n=13/23$), sometimes with no link between them beyond a shared module label, leaving readers to connect gene identity and spatial pattern on their own. \Cref{fig:coexpression_modules} shows a related variant of this pattern, juxtaposing a gene co-regulation network with its correlation matrix.

    \subsubsection{Functional Enrichment \& Pathway Activity}
    \label{sec:pathway}

    Once genes are ranked by spatial variability, a further question follows: what biological function do they serve? \taxtaskformat{Functional enrichment and pathway activity} analysis aggregates gene-level expression into pathway-level scores corresponding to biological processes such as oncogenic signaling, immune activation, metabolic programs, or the cell cycle, with each pathway represented as a set of tens to hundreds of member genes. Because the object of interest shifts from a single gene to sets of pathways, the visualization problem shifts from grounding a signal in tissue to comparing many scores against one another.
      
\begin{figure}[tbp]
      \centering
      \includegraphics[width=0.32\textwidth]{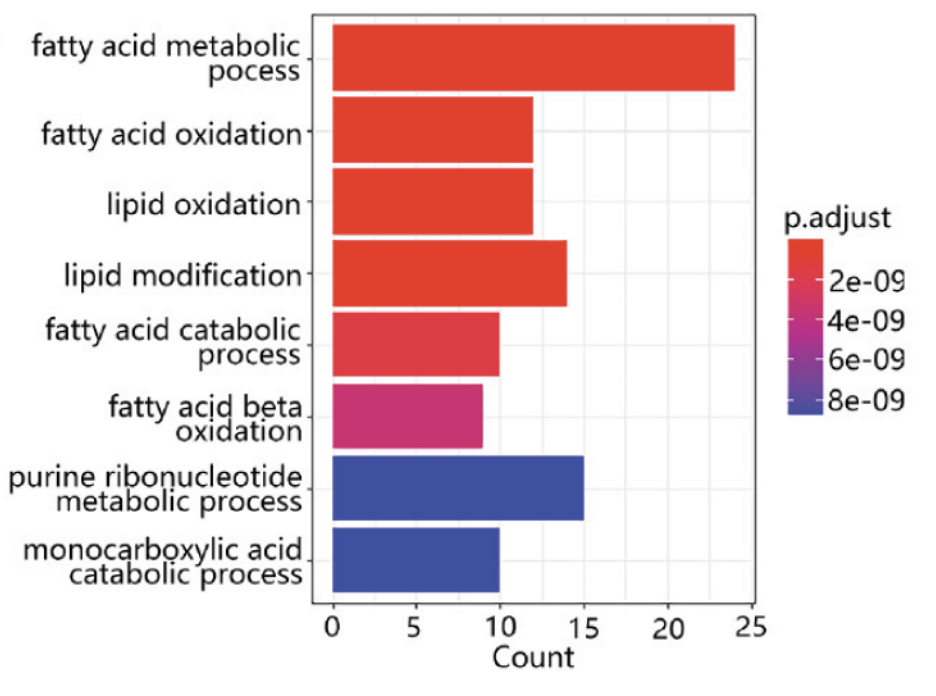}
      \caption{Example of \taxcodeformat{complete abstraction} in functional analysis, displaying a sorted linear \taxcode{bar-plot} of biological process enrichment scores ranked by significance. Reproduced from Feng \etal~\cite{Feng2025_1}. Panels adapted under CC BY 4.0.}
      \label{fig:pathway_enrichment_bar}
    \end{figure}

    This comparative framing is visible throughout how the figures are built. \taxcodeformat{linear, non-spatial} layouts dominate (85.1\%, $n=40/47$), led by \taxcodeplural{bar-plot} (44.7\%, $n=21/47$), and \taxcodeformat{complete abstraction} is applied in 89.4\% of papers ($n=42/47$). Enrichment scores and pathway rankings are shown as sorted \taxcodeplural{bar-plot} or \taxcodeplural{dot-plot} that make inter-pathway comparison efficient but discard spatial context entirely (see \Cref{fig:pathway_enrichment_bar}, for example). As in the previous sub-task, this trades one kind of interpretability for another. \taxcodeplural{spatial-map} of pathway scores, used in only 27.7\% of papers ($n=13/47$), can show \emph{where} a given pathway is active but not how it ranks against others; sorted \taxcodeplural{bar-plot} or \taxcodeplural{dot-plot} do the opposite, ranking pathways \emph{among} each other while discarding \emph{where} any one of them is active. In the majority of papers (78.7\%, $n=37/47$), \taxcode{sampling} is used to manage the large pathway sets typically evaluated, although selection criteria are often undisclosed.

    \subsection{Spatial Entity \& Domain Identification: Mapping Cellular and Structural Boundaries}
    \label{sec:spatial entity}
    
\begin{figure*}[t]
      \centering
      \includegraphics[width=\textwidth]{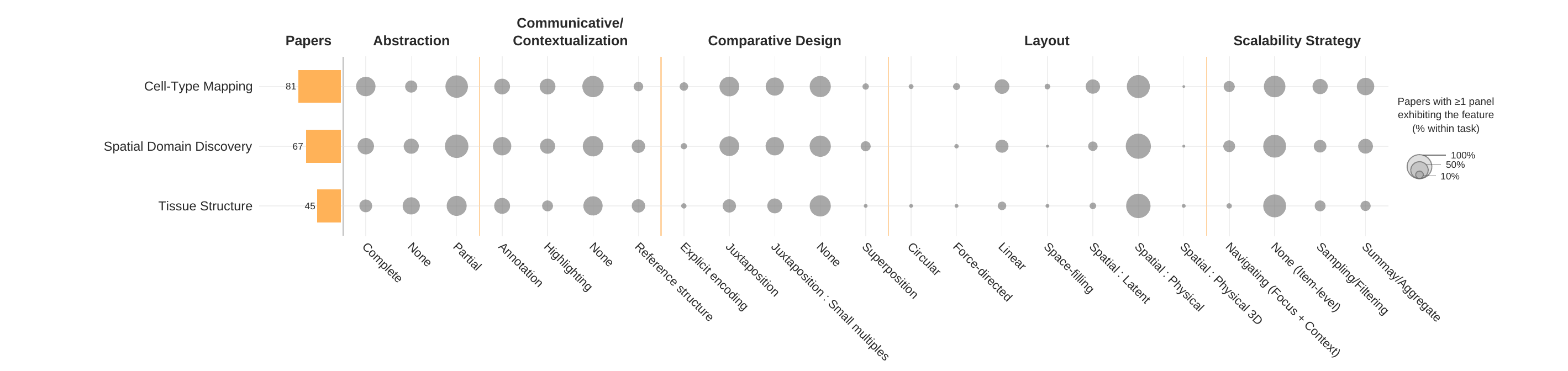}
    \caption{Visualization design patterns associated with \taxcode{task-spatial} tasks.}
    \label{fig:spatial_entity_identification}
    \end{figure*}

    While the previous section examined gene (co-)expression across space on a broader scale, the figures discussed in this section target the identification of more specific spatial structures. The corpus reflects this directly: spatial grounding stays close to its maximum across all three sub-tasks: (\textit{i})~\taxtask{cell-type}, (\textit{ii})~\taxtask{domain-discovery}, and (\textit{iii})~identification of \taxtask{tissue-structure}. \Cref{fig:spatial_entity_identification} summarizes the visualization design patterns observed across these three sub-tasks.
    
    \subsubsection{Cell-Type Mapping}
    \label{sec:cell-type}
    
    Assigning transcriptional identity to spatially indexed observations serves two demands: a \emph{confirmatory} demand, verifying that a known cell type occupies an anatomically expected location, and an \emph{exploratory} (or \emph{discovery}) demand, identifying \emph{de novo} spatial clustering or microenvironment-driven state shifts without a prior hypothesis of what will be found. We revisit this contrast between confirmation and exploration as a candidate for a cross-cutting axis of the \taxaxis{task} axis in \Cref{sec:discussion}. In our corpus, \taxtaskformat{cell-type mapping} is fundamentally an annotation task: \taxcode{annotations} appear in 85.2\% of papers ($n=69/81$), while the \taxcode{expression} serve mainly as a validation substrate.
    
\begin{figure}[b]
      \centering
      \includegraphics[width=0.45\textwidth]{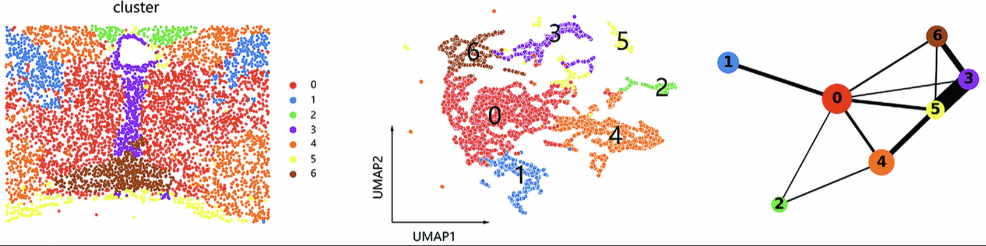}
      \caption{Example of a juxtaposed multi-view layout for cell-type mapping, displaying a spatial tissue cluster scatterplot (left), a corresponding non-spatial UMAP embedding (center), and a topological PAGA connectivity graph (right) sharing a unified categorical color encoding. Reproduced from Feng \etal~\cite{Feng2025_1}. Panels adapted under CC BY 4.0.}
      \label{fig:cell_type_triplet}
    \end{figure}

    To support both demands simultaneously, the field relies on a coordinated multi-view layout: the \taxcodeformat{physical spatial} layout is the majority idiom (79.0\%, $n=64/81$) and is frequently paired within the same paper with a non-spatial \taxcode{latent-layout}, such as one generated using UMAP or t-SNE (30.9\%, $n=25/81$). In \Cref{fig:cell_type_triplet}, for example, a spatial cluster scatterplot confirms where each cell type or state is located in the tissue, the latent embedding supports exploratory assessment of cluster separation, and the connectivity graph communicates inter-cluster relationships, all three sharing a single categorical color encoding.
    
    These idioms, however, break down as cell-type diversity grows: beyond roughly seven cell types, categorical colormaps lose perceptual separability and readability drops sharply~\cite{Giovannangeli2021, Tseng2023}. While designed to build trust and show consistency, this multi-view layout has two key limitations. First, categorical labels force a fixed annotation granularity upfront, collapsing the exploratory and confirmatory steps into a single static encoding. Second, automated cell-type annotations rarely visualize confidence scores, offering little support for validating the classification itself.

    \subsubsection{Spatial Domain Discovery}

    \label{sec:domains}

    Above the cell level, tissue organizes into regions with coherent functional \enquote{geography} (e.g., cortical layers, tumor regions, vascular niches). \taxtaskformat{Spatial domain discovery} infers these regions from expression data, often without a predefined reference of what the regions should be. Hence, the visualization must serve simultaneously as result display and a validation tool, regardless of whether the underlying method is unsupervised, supervised, or manually curated. This need for built-in validation shapes the data these figures draw on: \taxcode{native-2d} spatial coordinates are present in 95.5\% of papers ($n=64/67$) and \taxcode{annotations} in 92.5\% ($n=62/67$), while \taxcodeformat{gene expression matrices} appear in only 52.2\% ($n=35/67$). This indicates that domain figures typically display computational labels rather than the raw expression those labels were derived from.

\begin{figure}[tbp]
      \centering
      \includegraphics[width=0.45\textwidth]{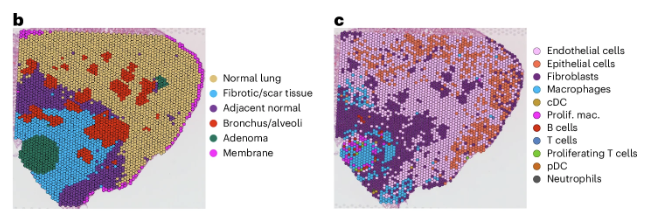}
      \caption{Example of juxtaposed partial abstractions for spatial domain validation. (b) Discrete grid-restricted hexagonal map displaying color-coded computational domain assignments. (c) Corresponding spatial map showing categorical cell-type distributions to visually reference and validate the computational boundaries. Reproduced from Aminu \etal~\cite{Aminu2025}. Panels adapted under CC BY 4.0.}\label{fig:spatial_domain_discovery}
    \end{figure}

    The same logic carries into layout choices. \taxcodeformat{physical spatial} layout dominates (94.0\%, $n=63/67$), the highest rate among entity/domain tasks, with \taxcodeplural{scatterplot} (50.8\%, $n=34/67$), \taxcodeplural{histology-image} (41.8\%, $n=28/67$), and \taxcodeplural{hexplot} (35.8\%, $n=24/67$) as the leading chart types. \taxcodeformat{partial abstraction} is used in 82.1\% of papers ($n=55/67$), preserving tissue outline and observation positions while replacing raw expression with discrete domain-membership colors. Validation, however, rarely stops at a single domain map: \taxcode{juxtaposition} (58.2\%, $n=39/67$) is widely used to place that map beside a histology image, cell-type map, or marker-expression map for cross-checking. \Cref{fig:spatial_domain_discovery} places a hexagonal domain assignment map alongside its corresponding cell-type map. \taxcode{references} such as anatomical outlines appear in 26.9\% of papers ($n=18/67$), one of the higher rates among all task categories, offering a further independent check against known anatomy.

    This way of validating domains, however, has a blind spot: discrete color labels communicate domain membership efficiently but obscure within-domain heterogeneity and impose false sharpness on what may be gradual transitions. Developmental gradients or invasion zones are rarely shown, and soft-assignment or probabilistic domain visualizations that could capture this uncertainty remain scarce in the corpus.

    \subsubsection{Tissue Structure}
    \label{sec:morphology}

    \taxtaskformat{Tissue structure} is the second most spatially grounded sub-task in the \taxcodeformat{spatial entity \& domain identification} task: \taxcodeformat{physical spatial} layout is used in 88.9\% of papers ($n=40/45$). Figures addressing this sub-task aim to establish macro-scale structural context, verify registration, or orient the reader, and are usually \taxcodeformat{juxtaposed} with complementary plots that provide more detailed information about gene expression, cell types, spatial domains, or other molecular and cellular patterns.

\begin{figure}[tbp]
      \centering
      \includegraphics[width=0.30\textwidth]{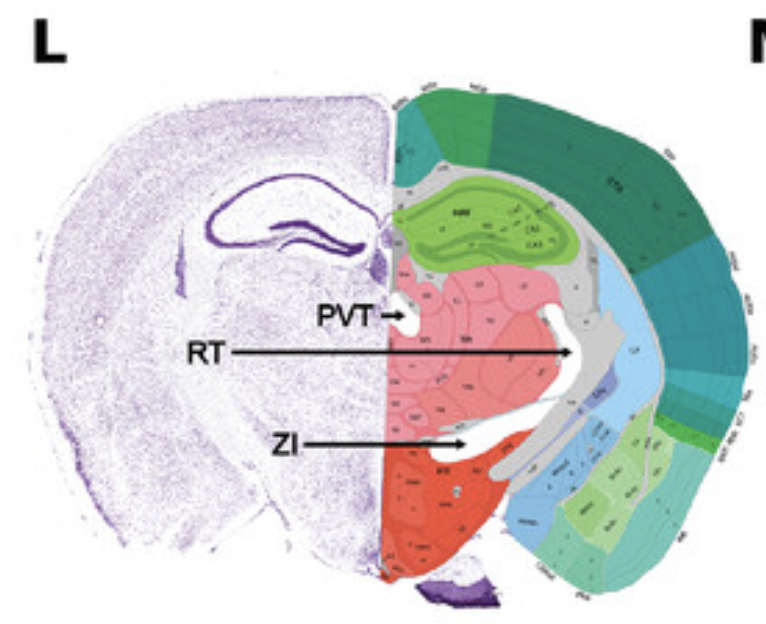}
      \caption{Example of a split-register morphological layout. 
          (Left) High-resolution histological tissue section (H\&E staining) preserving anatomical and structural realism. 
          (Right) Spatially registered anatomical reference atlas mapping regional boundaries and annotated functional geography across the identical physical substrate. Reproduced from Müller-Bötticher \etal~\cite{MllerBtticher2024}. Panels adapted under CC BY 4.0.}
    \label{fig:morphology_split_reference}
    \end{figure}

    \taxcodeplural{histology-image} (H\&E or immunofluorescence) are the leading chart type here (51.1\%, $n=23/45$). Although directly superimposing molecular markers onto a dense histological background is a common option, it tends to produce visual clutter. The dominant integration idiom is instead a split-anatomy layout: a high-resolution histological reference mirrored beside a spatially registered domain atlas or cell-type map. \Cref{fig:morphology_split_reference} shows this idiom applied to brain tissue, where a standardized anatomical atlas already defines the layers and regions of the reference panel. For tumor analysis, in contrast, boundaries typically separate healthy from diseased tissue. Therefore, instead of an atlas, these panels more often rely on direct annotations drawn onto the tissue itself. Such \taxcodeplural{annotation-how} were present in 37.8\% of papers in this sub-task ($n=17/45$). This split idiom preserves structural landmarks next to abstracted computational boundaries. While the morphological image often serves as an entry point to a multi-panel sequence, it is then abandoned in later, more analytical panels.

\subsection{Spatial Relational Networks: Topologies of Proximity and Interaction}
\label{sec:relational}

    \begin{figure*}[t]
      \centering
      \includegraphics[width=\textwidth]{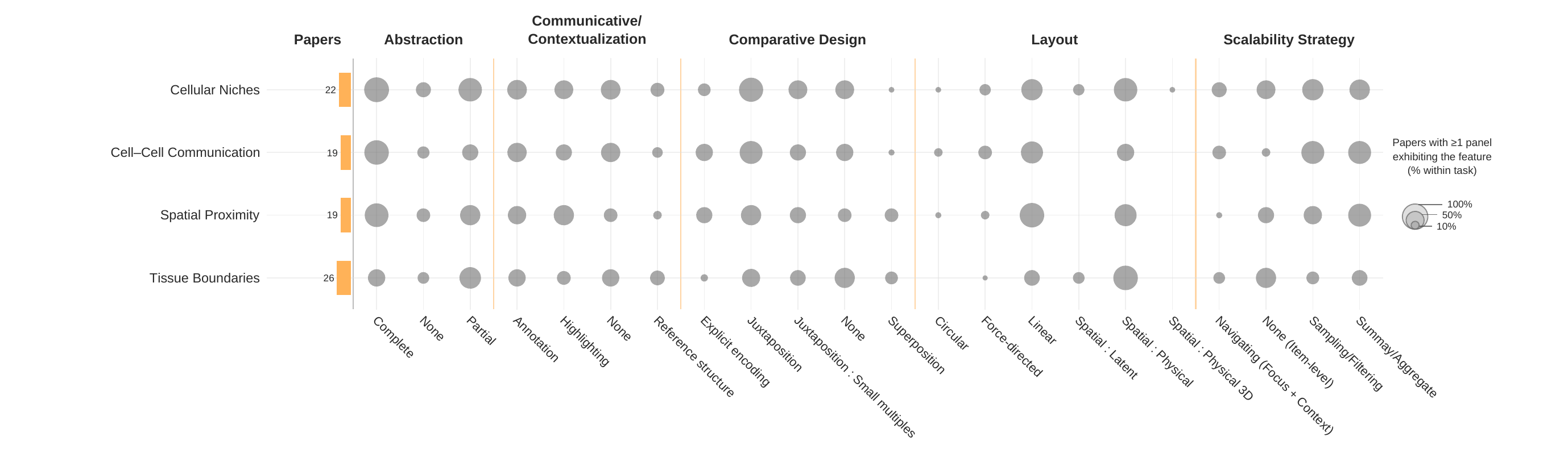}
    \caption{Visualization design patterns associated with \taxcode{task-relational} tasks.}
    \label{fig:spatial_relational}
    \end{figure*}

    The previous sections looked at spatial biological structure on a broader scale and from the perspective of individual entities (e.g., genes, cells, domains), asking how those entities are distributed or identified across tissue space. This section addresses a different form of biological reasoning: how entities relate to one another spatially, and what these relationships imply about biological function. Because the object of visualization is now a relationship between two or more entities, these tasks face a shared tension between preserving spatial fidelity and communicating relational topology. As \Cref{fig:spatial_relational} shows, however, this shared tension does not resolve into a single \enquote{relational} pattern. We identified sub-tasks with four different targets: (\textit{i})~\taxtask{cell-communication}, (\textit{ii})~\taxtask{cellular-niches}, (\textit{iii})~\taxtask{spatial-proximity}, and (\textit{iv})~\taxtask{tissue-boundaries}.

    \subsubsection{Cell--Cell Communication}
    \label{sec:ccc} 

    \taxtaskformat{Cell--cell communication} analysis infers molecular interactions between cell populations from expression patterns and ligand--receptor databases. \taxcodeformat{gene expression matrices} are ubiquitous data foundations for this inference (94.7\%, $n=18/19$), and are often augmented by \taxcode{annotations} (84.2\%, $n=16/19$). Spatial position, too, is often incorporated in the inference, constraining candidate interactions to spatially proximal cells or weighting interaction strength by distance. The underlying data substrate is therefore also strongly spatial: \taxcode{native-2d} coordinates are present in 84.2\% of papers ($n=16/19$). However, \taxcodeformat{physical} layout is used in only 42.1\% of papers ($n=8/19$), while non-spatial \taxcodeformat{linear} layouts dominate (68.4\%, $n=13/19$).

\begin{figure}[tbp]
        \centering
        \begin{minipage}[c]{0.55\columnwidth}
            \centering
            \includegraphics[width=\linewidth]
                {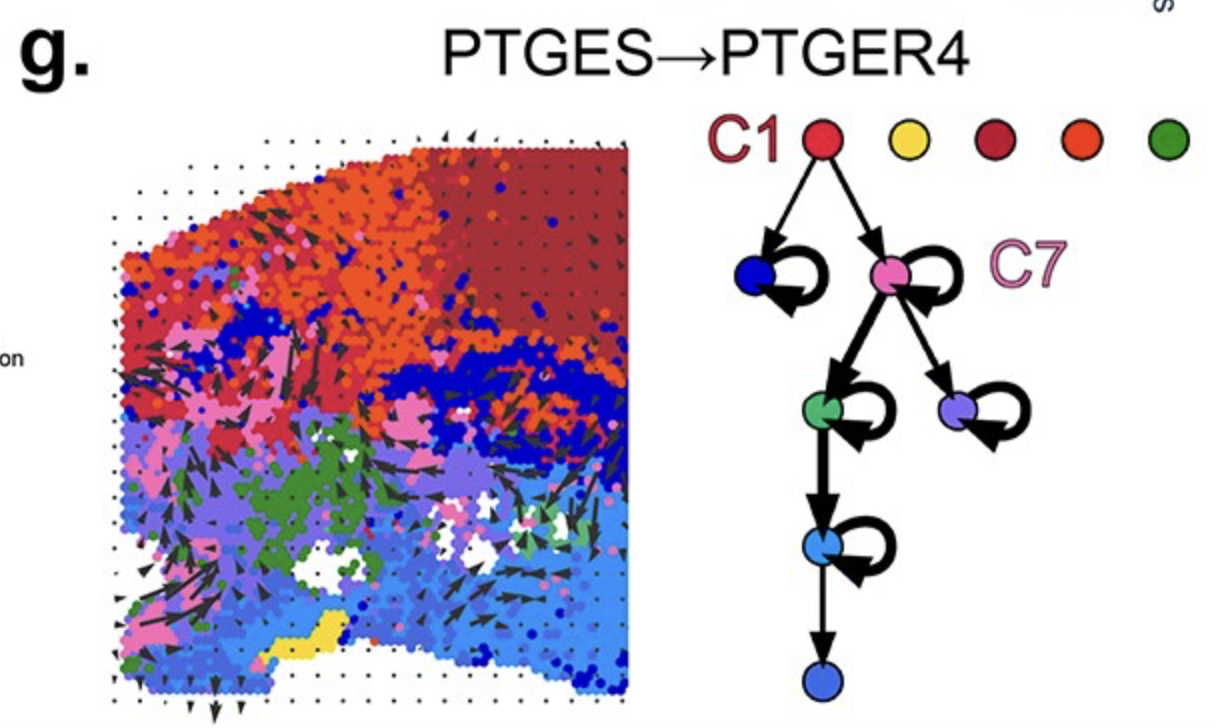}
        \end{minipage}
        \hfill
        \begin{minipage}[c]{0.40\columnwidth}
            \centering
            \includegraphics[width=\linewidth]
                {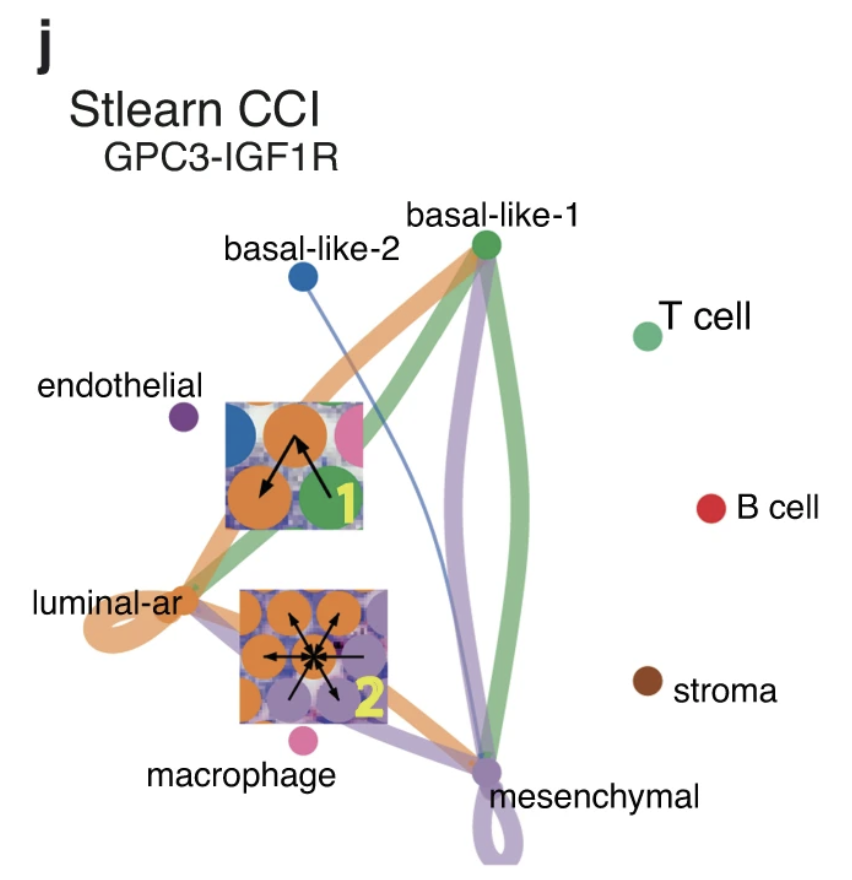}
        \end{minipage}
        \caption{Examples of complementary spatial and topological representations of cell signaling. Left: spatial tissue layout using a vector field to encode the direction and strength of interaction for a ligand--receptor pair (\textit{PTGES} $\rightarrow$
        \textit{PTGER4}), alongside a non-spatial directed graph summarizing signaling topology across cell clusters. Adapted from Pang
        \etal~\cite{Pang2025}. Right: abstract cell-type interaction network with physical focus-and-context anchoring, where localized
        high-resolution tissue insets validate spatial cell-to-cell contact.
        Adapted from Pham \etal~\cite{Pham2023}. Panels were cropped and
        combined; source materials licensed under CC BY 4.0.}
    
        \label{fig:ccc_combined}
    \end{figure}

    This tendency to abstract the space for this sub-task may arise from an encoding conflict. Abstract idioms such as \taxcodeplural{chord-diagram} and \taxcodeplural{network-graph} (each 21.1\%, $n=4/19$) answer \emph{what} interacts with \emph{what}, but not \emph{where}; physical maps answer \emph{where} a ligand or receptor is expressed but cannot represent the relationship itself. No canonical idiom bridges the two; instead the corpus shows several competing attempts to close that gap. Some papers pair a spatial vector field with an abstract directed graph, letting each chart type carry half the answer (see, e.g., \Cref{fig:ccc_combined} left). Others keep the abstract network as the primary layout and embed high-resolution tissue insets directly at its boundaries, anchoring individual edges back to physical contact (\Cref{fig:ccc_combined} right). And because the number of candidate interactions can be too large to encode statically at all, some authors choose to abandon fixed static idioms for interactive dashboards (see Supplemental Material Figure S6).

    \subsubsection{Cellular Niches}
    \label{sec:niches}
    
    \taxtaskformat{Cellular niches} are defined by spatial community structure, that is, repeating patterns of cell-type compositions across the tissue. This is distinct from spatial domains (defined by transcriptional identity) and ligand--receptor interactions (defined by signaling). Because the unit of interest is a neighborhood, visualization may encode aggregate, neighborhood-level properties rather than values for individual cells. For niche-related tasks, \taxcode{annotations} are the dominant data component (95.5\%, $n=21/22$), followed by \taxcode{native-2d} coordinates (90.9\%, $n=20/22$) and \taxcodeformat{gene expression matrices} (86.4\%, $n=19/22$).

    The neighborhood-level framing carries directly into layout choices. \taxcodeformat{physical spatial} layouts are markedly more common than \taxcodeformat{linear layouts} (77.3\% vs.\ 63.6\%, $n=17/22$ vs.\ $n=14/22$): niche papers typically pair a spatial scatterplot with a non-spatial compositional summary rather than abandoning spatial grounding altogether. \taxcodeplural{scatterplot} (50.0\%, $n=11/22$) and \taxcodeplural{heatmap} (40.9\%, $n=9/22$) are the most prevalent chart types. \taxcode{aggregation} is used in most papers to summarize neighborhood composition (59.1\%, $n=13/22$), and \taxcode{juxtaposition} is the dominant comparative strategy (81.8\%, $n=18/22$).

\begin{figure}[bt]
      \centering
      \includegraphics[width=0.45\columnwidth]{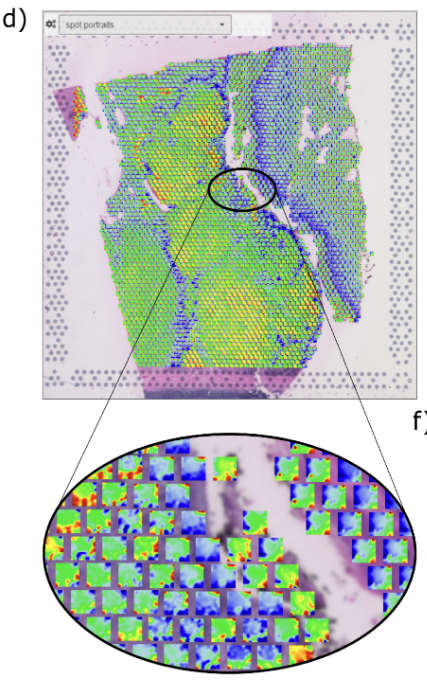}
      \caption{Example of a spatial layout for microenvironment visualization. Individual array spots embed aggregate multi-dimensional expression portraits to communicate localized molecular niche composition across physical tissue coordinates. Reproduced from Schmidt \etal~\cite{Schmidt2024}. Panels adapted under CC BY 4.0.}
      \label{fig:niche_som_interaction}
    \end{figure}

    However, pairing heatmaps with scatterplots remains a compromise: heatmaps summarize composition but lose spatial context, while spatial scatterplots show location but not composition. \Cref{fig:niche_som_interaction} bridges both views by placing a miniature SOM expression portrait at each spot coordinate. This preserves spatially resolved molecular profiles but requires interactive zoom and hover to manage visual density.

    \subsubsection{Spatial Proximity}
    \label{sec:proximity}
    
    \taxtaskformat{Spatial proximity} analysis ($n=19$ papers, 46 panels) tests whether cell types or features co-localize more than by chance, using spatial statistics such as distance distributions or cross-$K$ functions. The questions that figures within this sub-task aim to answer are closer to \enquote{is this significant} than \enquote{where does this occur,} and the layout choices follow this framing directly: \taxcodeformat{linear layouts} account for 84.2\% of papers ($n=16/19$), the highest rate among relational tasks, and \taxcode{aggregation} is the dominant scalability strategy (73.7\%, $n=14/19$).

\begin{figure}[bt]
      \centering
      \includegraphics[width=0.45\textwidth]{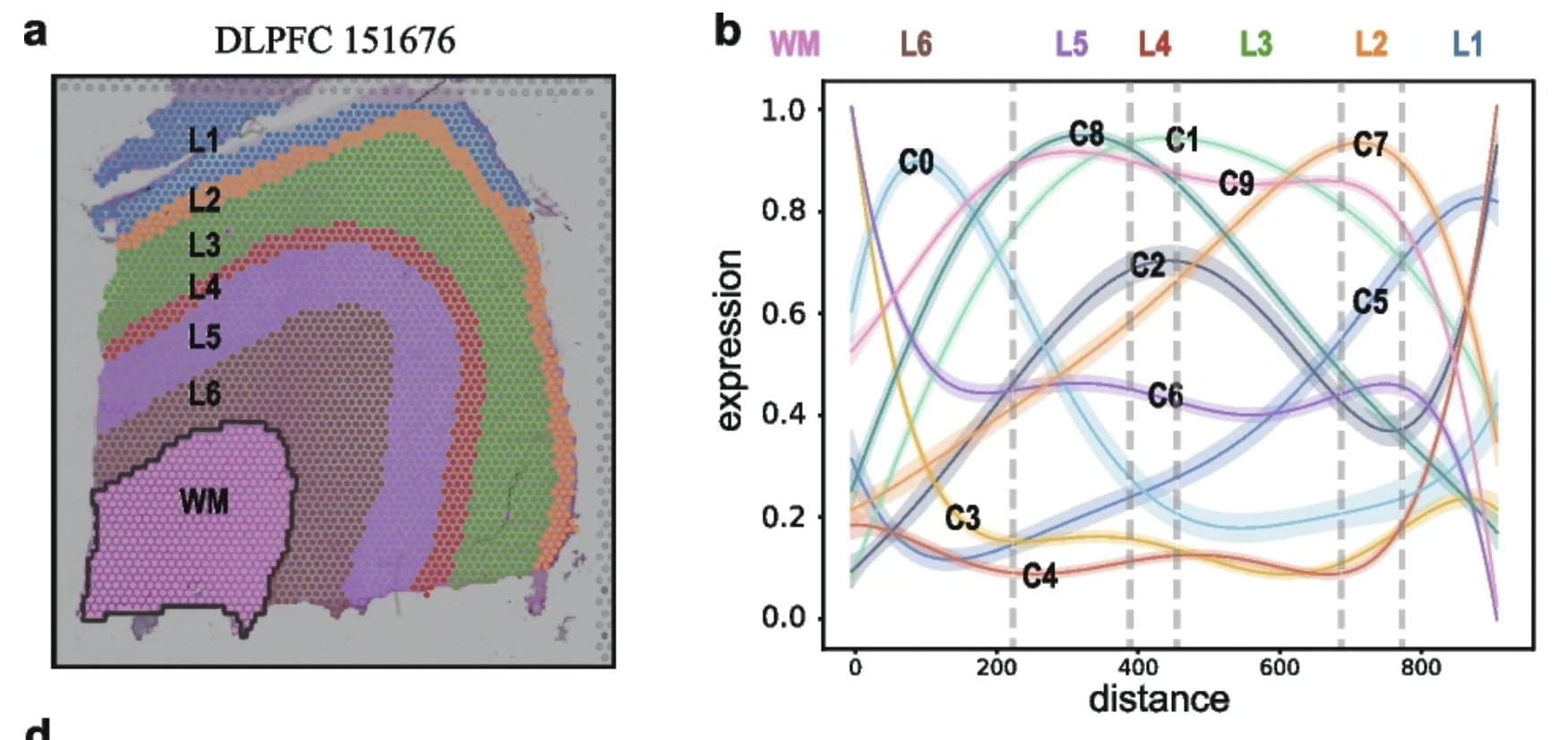}
      \caption{Example of juxtaposed morphological reference and complete spatial abstraction in proximity analysis. (a) High-resolution histological section of the human dorsolateral prefrontal cortex establishing native anatomical layer context and landmark boundaries. (b) Non-spatial linear regression curves modeling aggregated gene expression tendencies as a function of continuous geometric distance from the white matter boundary. Reproduced from Wang \etal~\cite{Wang2025_SOAPy}. Panels adapted under CC BY 4.0.}
      \label{fig:spatial_proximity_curves}
    \end{figure}

    \Cref{fig:spatial_proximity_curves} shows this pattern in practice: a histological reference establishes tissue context, then a companion panel abandons the coordinate system, plotting expression as a regression curve against distance from a boundary. This is efficient for testing significance, but a significant statistic computed over an entire section can still be driven by a small, spatially focal region of co-occurrence that the summary curve cannot reveal.

    \subsubsection{Tissue Boundaries}
    \label{sec:boundaries}

    A \taxtaskformat{tissue boundary}, a domain edge, or a tumor margin is a spatial object whose molecular gradient only makes sense in relation to where that boundary actually is. The data reflect this: \taxcodeformat{physical spatial} layout is used in 84.6\% of this sub-task's papers ($n=22/26$). Keeping the boundary anchored to space, while still showing how expression changes with distance from it, is usually solved with a two-step layout. The Matisse framework~\cite{MarcoSalas2021} shows one example of this approach: a line on the spatial map marks the boundary, while an adjacent chart plots expression against distance from that line. Unlike figures in cell-cell communication, this approach lets the reader trace the gradient back to a real location in the tissue, instead of reading it as just a number on a chart (see Supplemental Material Figure S7).
    

\subsection{Spatiotemporal Dynamics \& Perturbations: Visualizing Vectors and Transitions}
\label{sec:spatiotemporal}

The preceding sections examined figure panels addressing questions about biological structure as it exists at a given moment in time. This section turns to panels that address how biological states change: how cells progress through developmental or disease trajectories, and how these dynamics are distributed across tissue space. The two sub-tasks discussed in this section, (\textit{i})~\taxtask{lineage-trajectory} ($n=15$ papers) and (\textit{ii})~understanding \taxtask{evolving-processes} ($n=10$ papers), are the least frequent ones in the corpus. They reflect a domain in which analytical methods are still maturing. Despite their limited representation, these sub-tasks introduce distinct visualization challenges. As \ST moves beyond static description toward reconstructing dynamic biological processes, visualization must evolve accordingly, supporting reasoning not only about \emph{where} biological states occur, but also about \emph{how} and \emph{when} states transition, propagate, and reorganize across space and time.

\begin{figure*}[tbp]
  \centering
  \includegraphics[width=\textwidth]{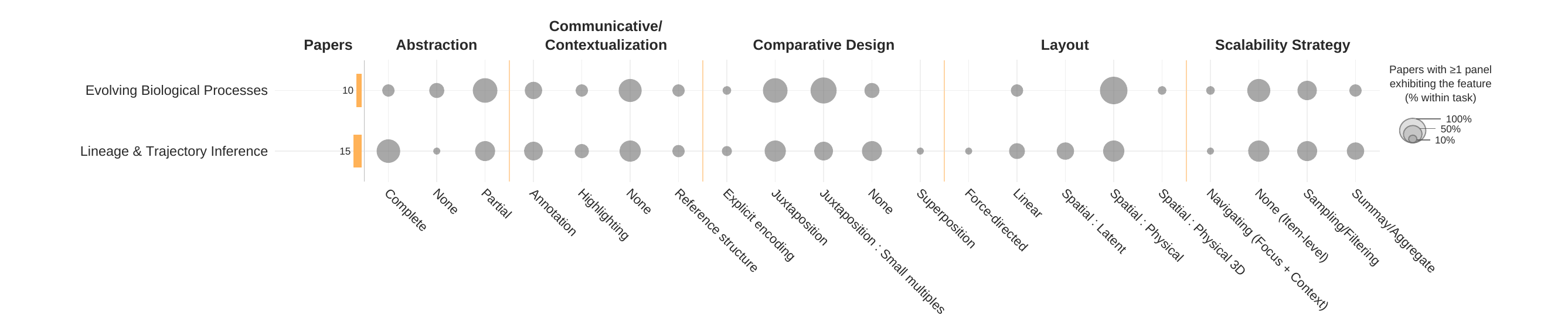}
\caption{\label{fig:patiotemporal_dynamics} Visualization design patterns associated with \taxcode{task-spatiotemporal} tasks.}
\end{figure*}

    \subsubsection{Lineage \& Trajectory Inference}
    \label{sec:trajectory}
    
\begin{figure}[tbp]
      \centering
      \includegraphics[width=0.75\columnwidth]{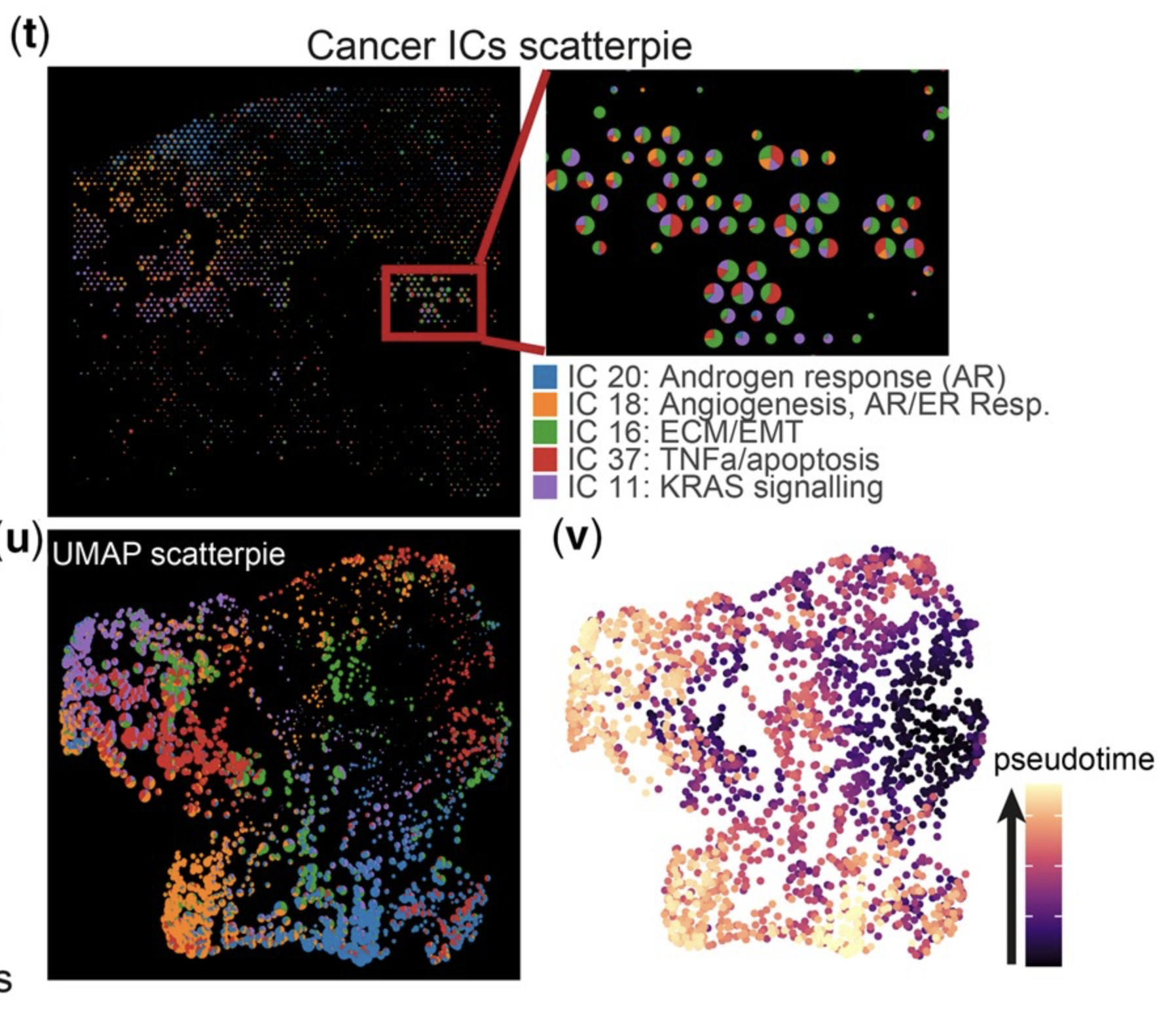}
      \caption{Example of juxtaposed latent trajectory topology and multi-dimensional spatial glyph layouts observed in our corpus. (v) Non-spatial latent UMAP projection encoding continuous pseudotime calculation to map trajectory paths and transitional cell states. Companion spatial (t) and latent UMAP (u) scatterpie representations mapping underlying independent component weights simultaneously across physical coordinates and cluster projections. Reproduced from Thuilliez \etal,~\cite{Thuilliez2024}. Panels adapted under CC BY 4.0.}
      \label{fig:trajectory_lineage_mapping}
    \end{figure}

    The figures that are part of this subset aim to answer two questions at once: which state does a cell transition to and where does that transition occur in the tissue. Consequently, this sub-task shows the smallest gap between \taxcodeformat{physical} and \taxcodeformat{latent} layouts of any entity-focused section in the corpus (60.0\% vs.\ 40.0\%, $n=9/15$ vs.\ $n=6/15$). This dual-layout pattern echoes the one seen in cell-type mapping, with a small difference: the \taxcodeformat{latent projection} encodes \emph{which} states transition to \emph{which}, at the cost of abstracting away from tissue geometry, while the spatial layout shows \emph{where} that transition unfolds. \taxcodeplural{scatterplot} (60.0\%, $n=9/15$) and \taxcodeplural{dimensionality-reduction} (40.0\%, $n=6/15$) are the most frequent chart types. To answer transition-related questions in static figures, researchers often pair two views: a latent UMAP embedding that captures similarities between cell states based on their gene expression profiles, with a spatially anchored view of the same transitional states, so that the topology of the transition and its physical location in the tissue can be both read from the figure. \Cref{fig:trajectory_lineage_mapping} is an example of this combination.

    The underlying data follow a similar pattern of dual representation: \taxcode{annotations} (80.0\%, $n=12/15$), \taxcodeformat{gene expression matrices} (73.3\%, $n=11/15$), and \taxcode{native-2d} coordinates (66.7\%, $n=10/15$) co-occur at rates broadly comparable to other sub-categories, while \taxcode{sequencing} (60\%, $n=9/15$) and \taxcode{imaging} (53.3\%, $n=8/15$) data are represented in near-equal proportion.

\subsubsection{Evolving Biological Processes}
\label{sec:evolution}

    Developmental morphogenesis, organ regeneration, and disease progression unfold over time, a dimension that static \ST can only partially access. This sub-category captures panels that reason about biological change over time, typically through multi-timepoint datasets. In the subset of figures corresponding to this sub-task, \taxcode{native-2d} coordinates are the most common data component (90.0\%, $n=9/10$), ahead of \taxcode{annotations} (70.0\%, $n=7/10$). \taxcodeformat{physical spatial} layout is used in \emph{every} paper in this sub-task (100\%, $n=10/10$). A~minority of papers additionally use 3D \taxcodeformat{physical} layouts (10.0\%, $n=1/10$) and \taxcode{reconstructed-3d} coordinates (20.0\%, $n=2/10$), reflecting the analysis of reconstructed tissues across different time points. 

\begin{figure}[tbp]
      \centering
      \includegraphics[width=0.45\textwidth]{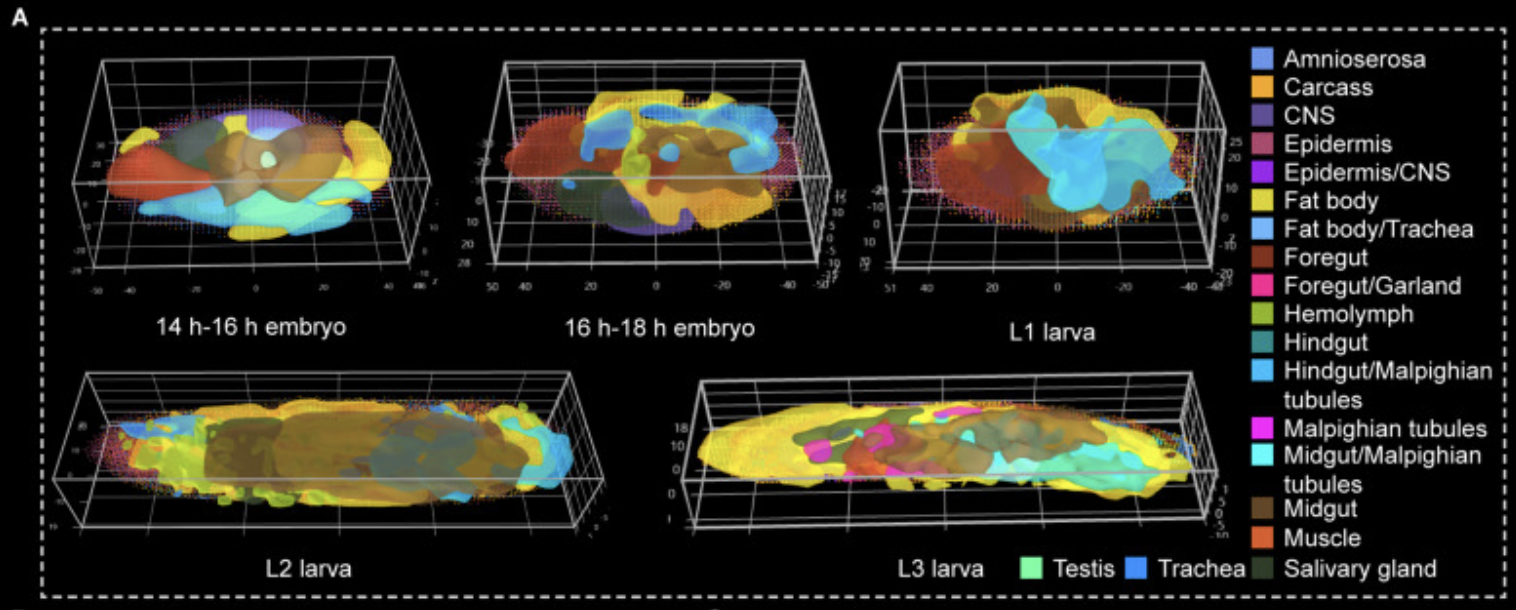}
      \caption{Example of juxtaposed 3D physical layouts for longitudinal developmental analysis observed in our corpus. Spatiotemporal atlas of Drosophila embryonic and larval development reconstructed in 3D space from high-resolution Stereo-seq profiles. Succession illustrates both the representation of volumetric continuity and the inherent challenges of structural occlusion in static 3D figures. Reproduced from Guo \etal~\cite{Guo2023}. Panels adapted under CC BY 4.0} 
          \label{fig:evolution_3d_morphogenesis}
    \end{figure}

    \taxcodeformat{Small-multiple juxtaposition} (90.0\%, $n=9/10$) is the dominant comparative strategy, commonly placing successive developmental or disease stages side by side, and \taxcodeformat{partial abstraction} is especially common (80.0\%, $n=8/10$). Juxtaposed tissue maps make temporal change easy to compare but require readers to mentally register regions across samples as morphology changes. This is even more challenging for reconstruction-based figures, which preserve volumetric continuity at the cost of occlusion and depth ambiguity. The longitudinal 3D Drosophila organogenesis atlas (see \Cref{fig:evolution_3d_morphogenesis}) is an example of juxtaposed 3D layouts that reveal how different tissue types evolve over time and space.

\subsection{Comparative \& Multi-Dataset Integration: Alignment, Scaling, and Resolution}
\label{sec:comparative}

\begin{figure*}[t]
  \centering
  \includegraphics[width=\textwidth]{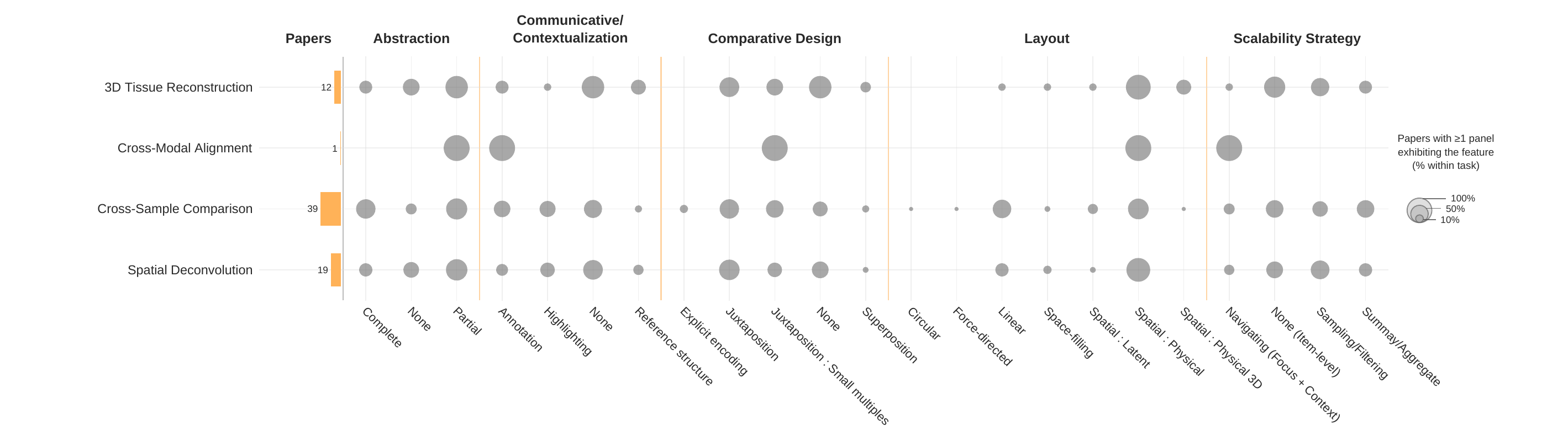}
\caption{ Visualization design patterns associated with \taxcode{task-comparative} tasks.}
\label{fig:comparative_multi}
\end{figure*}

Scientific knowledge in \ST accumulates by comparing findings across samples, conditions, modalities, and scales. This section covers how analysts visually communicate such comparisons, bridge technologies, or reconstruct 3D volumes from serial sections, largely by adapting visualization conventions from non-spatial contexts. \Cref{fig:comparative_multi} summarizes the visualization design patterns observed across the four sub-tasks: (\textit{i})~\taxtask{cross-sample}, (\textit{ii})~\taxtask{cross-modal}, (\textit{iii})~\taxtask{deconvolution}, and (\textit{iv})~\taxtask{reconstruction-3d}.

    \subsubsection{Cross-Sample Comparison}
    \label{sec:crosssample}
    
    The figures corresponding to this sub-task ($n=39$ papers, 185 panels) aim to answer whether a spatial pattern observed in one sample holds or changes across donors, disease states, conditions, or other researcher-defined axes of comparison. \taxcode{annotations} (71.8\%, $n=28/39$), \taxcodeformat{gene expression matrices} (69.2\%, $n=27/39$), and \taxcode{native-2d} coordinates (69.2\%, $n=27/39$) are the primary data components encoded in these figures, with \taxcode{sequencing} data clearly dominating over \taxcode{imaging} data (61.5\% vs.\ 38.5\%, $n=24/39$ vs.\ $n=15/39$).
    
\begin{figure}[bt]
    \centering

    \begin{minipage}[c]{0.48\columnwidth}
        \centering
        \includegraphics[width=\linewidth]
            {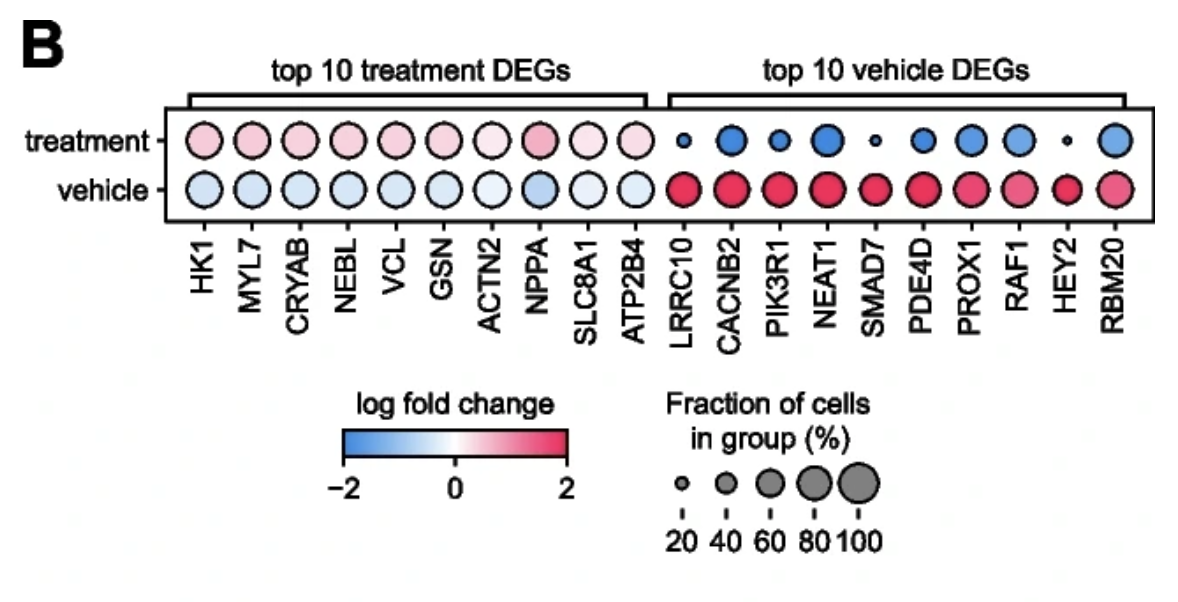}
    \end{minipage}
    \hspace{0.01\columnwidth}
    \begin{minipage}[c]{0.48\columnwidth}
        \centering
        \includegraphics[width=\linewidth]
            {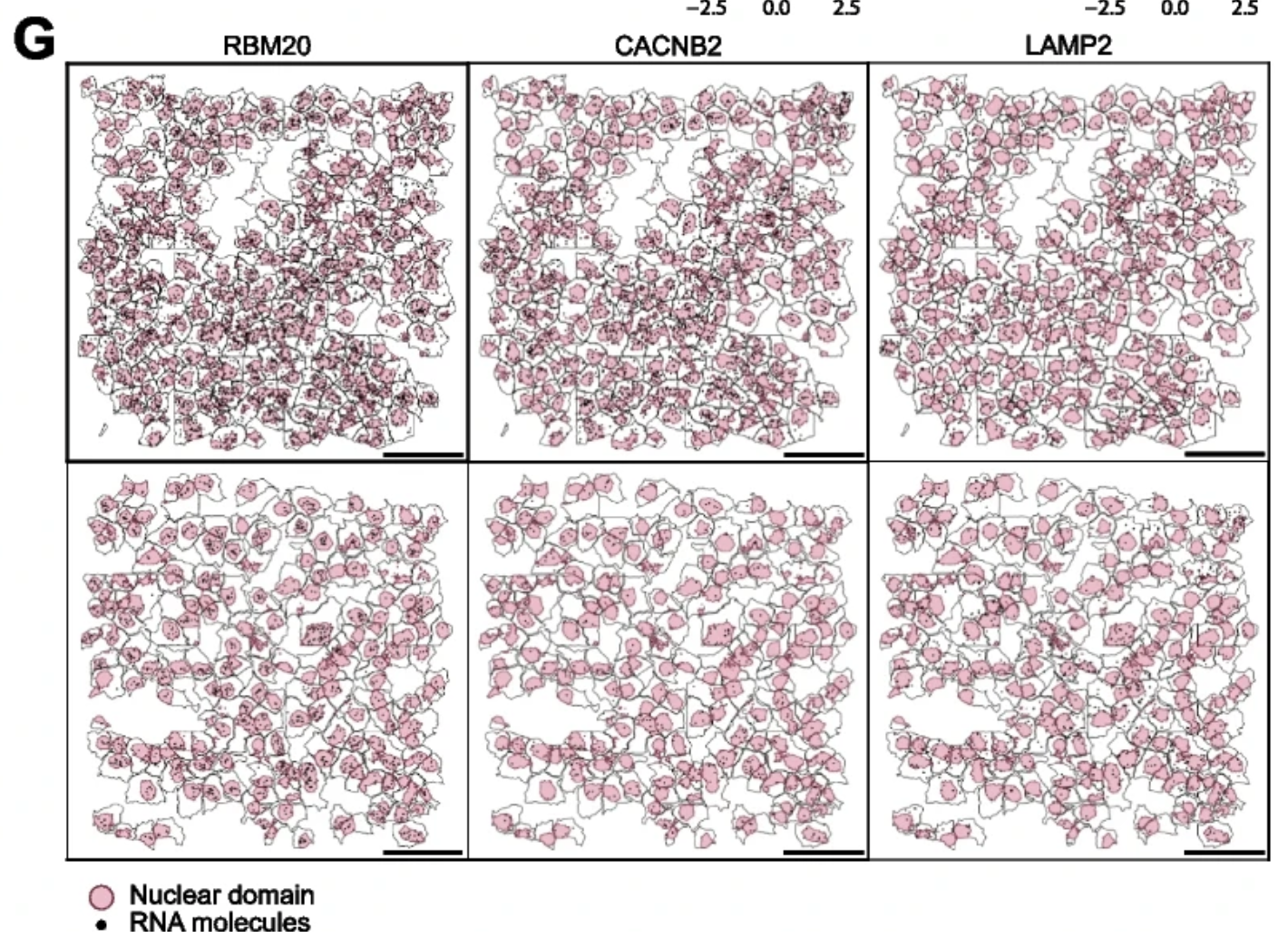}
    \end{minipage}

    \caption{Example of juxtaposed complete abstraction and physical small
    multiples for cross-condition comparison. Left (original panel b):
    completely abstract statistical representation mapping differential gene
    expression shifts between the vehicle and treated cell populations.
    Right (original panel g): companion physical spatial layout utilizing
    small-multiple juxtaposition to visually confirm condition-specific
    transcript redistribution across cellular coordinates. Adapted from Mah
    \etal~\cite{Mah2024}. Panels were cropped and combined; source material
    licensed under CC BY 4.0.}

    \label{fig:cross_condition_subcellular}
\end{figure}

    The emphasis on summarized evidence directly shapes figure design. \taxcodeplural{scatterplot} are the most commonly used chart type (51.3\%, $n=20/39$), but unlike in earlier sections, here they are predominantly non-spatial: \taxcodeformat{linear non-spatial} layouts (51.3\%, $n=20/39$) and \taxcodeformat{complete abstraction} (56.4\%, $n=22/39$) are more prevalent here than in the other integrative sub-tasks, suggesting that cross-sample reasoning tends to operate on summarized rather than spatially resolved data. \taxcode{item-level} and \taxcode{aggregation} are equally common scalability strategies (46.2\% each, $n=18/39$), and \taxcode{juxtaposition} is the primary comparative design (56.4\%, $n=22/39$). \Cref{fig:cross_condition_subcellular} captures the use of varying levels of abstraction: a completely abstract statistical panel isolates differentially expressed genes through a linear layout that prioritizes significance over space; it is juxtaposed with a physical layout that uses small multiples of raw transcript coordinates to visually confirm condition-specific transcript redistribution across the tissue.

\subsubsection{Cross-Modal Alignment}
\label{sec:crossmodal}

With only one qualifying paper in this sub-category ($n=1$ paper, $6$ figure panels), \taxtaskformat{cross-modal alignment} is the least-represented task in the corpus. The paper uses a \taxcodeformat{physical spatial} layout with \taxcodeformat{partial abstraction}, aligning with broader single-cell atlas literature, where multimodal data is most commonly displayed as parallel embedding plots~\cite{Cao2022, Hao2021}. Rather than trying to infer a general pattern from a single publication, we flag cross-modal spatial alignment as an open design gap. Given the biological importance of jointly visualizing transcriptomics, proteomics, and imaging, we see this gap as a priority for future work.


\subsubsection{Spatial Deconvolution}
\label{sec:deconvolution}

\taxtaskformat{Spatial deconvolution} outputs a cell-type proportion vector, rather than direct measurements. Even so, the task stays strongly spatially grounded to address \emph{where} specific cell types reside in low-resolution spots. This spatial grounding shows up clearly in the data: \taxcode{native-2d} coordinates appear in 94.7\% of papers ($n=18/19$), and \taxcodeformat{physical spatial} layouts are used in 84.2\% ($n=16/19$). These often manifest as \taxcodeplural{scatterplot} (52.6\%, $n=10/19$) or \taxcodeplural{hexplot} (26.3\%, $n=5/19$). \taxcodeformat{partial abstraction} appears in 68.4\% of papers ($n=13/19$), since deconvolution replaces raw expression values with proportion vectors as the entity being encoded.

\begin{figure}[bt]
    \centering    
    \vspace{2mm}
    \includegraphics[width=0.4\columnwidth]{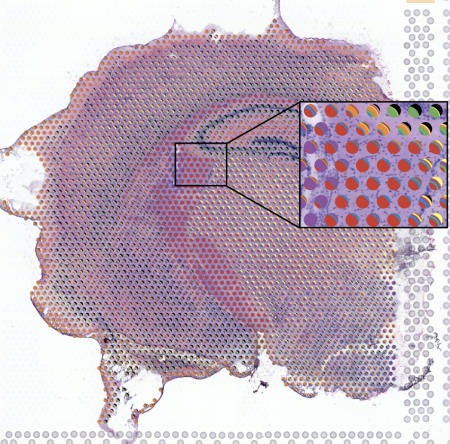}
    \caption{Example of a compact part-to-whole visualization in which proportions are represented as angular slices within a circular glyph. Reproduced from Zhao \& Marai~\cite{Zhao2024_preprint}. Panels adapted with permission from authors.}
    \label{fig:deconvolution}
\end{figure}

We found two specific encoding strategies to be the most common: coloring each spot by the proportion of one cell type using a sequential colormap, or using a small pie glyph per spot, often placed over an H\&E image. Each approach has its trade-offs. Coloring by a single cell type is easy to read but only shows one type at a time. Pie glyphs show the full mix of cell types but get hard to read as spots get denser. In fact, no idiom in the corpus shows more than one or two dimensions of the full composition at once. \Cref{fig:deconvolution} shows one alternative: a circular glyph that uses a slicing scheme that differs from traditional pie charts in an attempt to make multi-part compositions easier to read.

\subsubsection{3D Tissue Reconstruction}
\label{sec:registration}

Figures in this sub-task aim to answer how a spatial pattern holds across 3D tissue. Given the technical limitations and high cost of directly measuring the transcriptome of 3D tissue, these figures instead visualize the reconstruction of 3D tissue through computational frameworks~\cite{Lin2025}. Regarding the data, \taxcode{annotations} (83.3\%, $n=10/12$), \taxcode{native-2d} coordinates (50.0\%, $n=6/12$), and \taxcodeformat{gene expression matrices} (50.0\%, $n=6/12$) are the primary data components.  As expected, \taxcode{reconstructed-3d} coordinates appear in 58.3\% of papers ($n=7/12$) and \taxcode{true-3d} coordinates in 33.3\% ($n=4/12$). Several papers pair \taxcode{native-2d} with \taxcode{reconstructed-3d} to display input sections or validation views; others rely on 2D projections rather than explicit volumetric encodings, explaining why 3D coordinate codes are not universal. \taxcode{imaging} (58.3\%, $n=7/12$) and \taxcode{sequencing} (33.3\%, $n=4/12$) data are both represented.

\taxcodeformat{physical spatial} layout (91.7\%, $n=11/12$) is the most frequently used layout here. The 3D \taxcode{scatterplot} remains the single most frequent chart type (66.7\%, $n=8/12$), despite its well-known shortcomings: loss of depth information, problems caused by occlusion, and the inability to resolve these issues in non-interactive, static figures. \taxcodeplural{histology-image} rank second (16.7\%, $n=2/12$). The large majority of papers \taxcodeformat{do not make use of comparison} (75\%, $n=9/12$), reflecting that these figures primarily demonstrate feasibility rather than compare biological conditions. This pattern is illustrated in Figure S8: a single 2D reference slice is used to computationally reconstruct the tissue in three dimensions, and the same pipeline scales to combine multiple serial sections into one continuous 3D volume---without introducing comparison across biological conditions or donors.

\subsection{Interaction Support: Exploration and Its Limits}

The previous sections characterize how \ST findings are communicated through static publication figures. A~complementary picture emerges from the interactive systems in our corpus: 32 of the analyzed papers describe visualization tools, analysis platforms, or data frameworks with interactive visualization components. Because interactivity can rarely be inferred from individual static figures, we coded these systems at the paper or tool level. We used the taxonomy established by Yi \etal~\cite{Yi2007}, which distinguishes seven categories of interaction intent: \taxcode{explore}, \taxcode{select}, \taxcode{filter}, \taxcode{encode}, \taxcode{reconfigure}, \taxcode{connect}, and \taxcode{abstract-elaborate}.

Across the tools, \taxcode{explore} (31 tools) and \taxcode{filter} (31 tools) are the most consistently supported intents, followed closely by \taxcode{encode} (28 tools). \taxcode{reconfigure} (22) and \taxcode{connect} (21) appear in roughly two-thirds of tools, while \taxcode{select} (19) and, especially, \taxcode{abstract-elaborate} (13) are the least common (\Cref{fig:interactivity_taxonomy}). This distribution suggests that most \ST tools are built around an exploratory-inspection loop: users pan, zoom, and navigate tissue space (\taxcode{explore}), restrict the view by gene, cell type, sample, or region (\taxcode{filter}), and switch between alternative representations such as spatial maps, embeddings, heatmaps, and metadata views (\taxcode{encode}). The prevalence of \taxcode{connect} confirms that most systems operate as coordinated multi-view environments, an interactive counterpart to the multiple static layouts already documented in \Cref{sec:crossmodal,sec:deconvolution}.

Tools such as Vitessce~\cite{Keller2024} make this coordination concrete, directly operationalizing the multiple layouts described for \taxtask{cell-type} (\Cref{sec:cell-type}): a spatial \taxcode{scatterplot}, a UMAP embedding, and a \taxcode{heatmap} are shown side by side, and selecting cells in one view highlights them in every other view (\taxcode{connect}), while widgets let users retarget color encodings (\taxcode{encode}) and restrict the displayed population (\taxcode{filter}). Other tools such as Giotto~\cite{Dries2021}, Squidpy~\cite{Palla2022}, and TissUUmaps 3 ~\cite{Pielawski2023} follow the same pattern. Where \Cref{sec:crossmodal} noted that static categorical color breaks down past many cell types, this interactive linking offers a partial remedy: analysts can isolate subsets dynamically rather than relying on a single fixed legend.

This coordination, however, rarely extends to changing the level of detail within a single view. Of 21 tools that support \taxcode{connect}, only 11 (52.4\%) also support \taxcode{abstract-elaborate}: coordination across views does not reliably extend to elaboration within them. Few tools support the kind of continuous elaboration needed to resolve the same intermediate-representation gap identified for \taxtask{cross-sample} (\Cref{sec:crosssample}) and \taxtask{deconvolution} (\Cref{sec:deconvolution}), where static figures were forced to choose between one coarse encoding or an illegible fully elaborated one.  These limits in interactive elaboration echo the static figure patterns described above, and together they motivate the broader challenges and design opportunities we turn to next.

\section{Challenges and Opportunities}
\label{sec:challenges and opportunities}

\ST is a rapidly advancing field. In less than a decade, technologies have progressed from targeted gene expression panels to whole-transcriptome profiling, from multi-cellular capture spots to sub-cellular resolution, and from small 2D fields-of-view to higher-throughput 2D and 3D technologies~\cite{Lim2025, Moses2022}. Each technological leap expands what biological phenomena can be measured, and in doing so, sharpens the demands placed on visualization. The challenges documented in \Cref{sec:results} are a moving target driven by a persistent friction between advancing measurement technologies and lagging visual conventions.

The most immediate pressure point is spatial resolution. As data transitions from multi-cell spots to thousands of single-cell coordinates, discrete spatial glyphs that communicated expression cleanly at coarser resolutions dissolve into unscalable, overplotted point clouds. Furthermore, cell segmentation should not be considered a solved preprocessing step; its unsolved nature presents important visualization challenges~\cite{Ishaque2026}. With up to 40\% of transcripts unassigned to cells, visualization tools must shift away from deterministic cell-boundary polygons toward uncertainty-aware strategies~\cite{Salas2025, Ergen2025}.

This resolution increment generates a second challenge: the gap between molecular breadth and spatial legibility. The field's standard response to high-dimensional features remains the small-multiple tissue map, one facet per gene. While effective for select canonical markers, this strategy breaks down as a discovery interface at whole-transcriptome scale. No established visual idioms allow analysts to efficiently survey thousands of spatially variable features to discover co-expression modules or evaluate pathway boundaries directly within tissue space. Scalable overview representations that adapt dimensionality reduction principles to physical, tissue-embedded constraints remain a high-value design target. 

As a consequence of this dimensionality, the pressure towards abstraction frequently results in a total loss of spatial grounding. As biological reasoning shifts from localization toward functional interpretation, physical tissue geometry recedes from visual encodings. While each abstraction may be rational on its own, their cumulative effect can isolate vital findings---such as spatial zones of tumor infiltration or localized signaling programs---from the anatomical structures that give them meaning. Side-by-side coordinated views remain a common strategy for linking these representations, though further work is needed to understand how well this decoupled arrangement supports tasks that benefit from physical spatial coordinates. Exploring alternative designs, such as unified idioms where tissue geometry acts as a persistent, structural frame for abstract functional data, may be a promising direction.

Finally, technological progress has dramatically shifted the scale of comparative analysis. While early \ST workflows compared two or three tissue slices, current studies analyze dozens of specimens, pointing toward a clinical scale that strains traditional visualization designs. The complexity increases further when considering specimens from multiple conditions, such as different experimental designs, computational parameters, disease subtypes, or perturbations. This scalability challenge for small-multiple displays has already been recognized in the broader visualization literature, both as a design problem and a perceptual one. For design, prior work has developed formal models for structuring comparisons across many categories in small-multiple displays~\cite{Kehrer2013}. For perception, there is empirical evidence that comparison accuracy declines as the number of small-multiple frames increases~\cite{Hosseinpour2025}. Consistently, small-multiple layouts become less frequent as specimen or condition counts increase, well before typical cohort sizes are reached. To achieve scalability, current tools collapse spatial structures into non-spatial cell-type proportion charts, discarding the geographic patterns central to the comparison. \ST visualization requires intermediate spatial summary representations, encodings that preserve key geometric boundaries and variations that can scale to many specimens, conditions, or otherwise orthogonal variables. Addressing these cross-cutting bottlenecks of resolution, abstraction, multiscale integration, and uncertainty represents a profound opportunity to shape the visual language of spatial biology.


\section{Discussion}
\label{sec:discussion}

\ST reconnects molecules with tissue structure, but its visual language has yet to consistently reflect that connection. This study examined how published \ST figures represent biological data and support analytical reasoning. The central finding is that spatial faithfulness in published figures is strongly task-dependent: visualization practice is spatially grounded when biological questions concern entity identity and location, but systematically abandons spatial structure as reasoning moves toward functional interpretation and relational analysis. This finding reflects a mismatch between the complexity of the biological questions being asked and the visual conventions currently available to answer them.

This tension becomes concrete when tasks are examined in sequence. Cell type mapping and spatial domain discovery predominantly use tissue-preserving layouts. As biological reasoning scales toward pathway enrichment and cell--cell communication, physical spatial structure recedes in favor of abstract representations, like UMAP embeddings, chord diagrams, and enrichment bar charts. Each convention solves a real problem within its own analytical context, but all were developed for non-spatial data and imported without redesign. 

Their connection back to physical tissue is typically mediated by a shared color palette between an abstract chart and a discrete spatial map, which establishes categorical reference but cannot encode continuous geometry or physical proximity. This reliance on color as the only linking channel also inherits the well-documented perceptual limitations of color encodings: discriminability degrades sharply beyond seven categories, and correspondences become unreliable for readers with color vision deficiencies, compounding the loss of spatial information with a loss of categorical precision~\cite{Giovannangeli2021,munzner2015visualization, Wong2011, Ware2012}. In many cases, readers must reconstruct spatial reasoning that figures do not provide. The field has evolved its visual conventions organically over time, borrowing from single-cell atlases, bulk RNA-seq pipelines, and ligand-receptor tools, without a systematic framework to assess whether those conventions remain appropriate for tasks in a spatial setting.

This pattern of borrowing without redesign produces three recurring and consequential failures. First, functional and relational analyses lose spatial grounding. Most of the cell--cell communication figures routinely show interaction strength and directionality~\cite{Armingol2020} but not whether (or to what extent) communicating populations are physically co-localized in the tissue. What gets lost is precisely the information that distinguishes \ST from earlier single-cell methods. Second, uncertainty is present at multiple stages of the typical analysis pipeline: domain boundary assignment, cell-type deconvolution, and ligand-receptor inference are just a few examples of intrinsically uncertain processes. However, the publication figures we examined almost universally render their outputs as sharp, deterministic categorical boundaries. When quantitative uncertainty estimates are available from the underlying computational methods, they are rarely carried through into the figure. We worry that this practice may overstate analytical confidence and encourage overinterpretation in biologically ambiguous regions. While we  urge future authors to encode any available uncertainty information in their visualizations, we understand that this represents a major challenge in many cases. Third, cross-sample comparison has no viable intermediate representation, or at least none yet adapted for this domain. Small-multiple tissue maps preserve geometry but scale poorly, whereas aggregated summaries scale well but discard spatial structure entirely. As \ST studies grow toward clinical cohort sizes, this visual bottleneck will intensify. 

Identifying these gaps precisely required an analytical approach that existing reviews lack. Unlike traditional tool surveys or chart-type catalogs, this study contributes a systematic analysis of how a scientific field translates biological questions into visual form. Our task coding makes explicit the relationship between what is being asked biologically and how the visualization responds. This task-driven foundation, in turn, points toward specific and targeted design interventions for the visualization community. 

We argue that future visualization research should prioritize three concrete technical targets. First, functional and relational tasks need idioms that maintain spatial grounding while representing high-dimensional relational structure. Potential ideas include spatial network layouts that encode physical co-localization alongside interaction topology, or spatially resolved pathway summaries that show where activity is distributed across morphological context. Second, uncertainty visualization represents a clear and tractable opportunity. Paradigms like value-by-alpha maps, probabilistic contours, and soft categorical blends could be directly adapted and evaluated for domain boundaries and deconvolution outputs, building on established general-purpose frameworks~\cite{Correl2015, Roth2010, Correll2018} that have not, to our knowledge, been adapted for \ST. Third, cross-sample comparison requires new intermediate representations that preserve geometric variation across specimens without demanding full, unscalable spatial detail for each. Beyond individual idioms, future software systems must address a systemic limitation: current tools support exploratory inspection effectively but are consistently weaker at carrying analytical context forward, that is, preserving the spatial context and provenance of a finding as it moves across linked views, scales, and derived representations. This distinction echoes the broader visual analytics literature on analytic provenance~\cite{Xu2020}, although its application to spatial-omics tooling specifically appears to be a genuinely open question.

Finally, it is important to acknowledge what the present study can and cannot establish. Because we primarily analyzed static publication figures, our findings describe settled communicative choices rather than the full range of live analytical practice. Interactive workflows may address some of the gaps documented here, without ever appearing as static figure panels in print. Similarly, the panel-level unit of analysis gains systematic scale at the cost of sensitivity to multi-panel compositional arguments, and the frequency of a visualization type does not directly measure its perceptual effectiveness or impact on biological inference. The corpus is also weighted toward computational methods papers, meaning clinical application literature, with different communicative goals and audiences, may exhibit different patterns. These constraints define the scope of the findings: publication figures are precisely where visual conventions are fixed, peer-reviewed, and propagated, and characterizing them systematically is itself a tractable and previously unavailable empirical contribution. In consequence, our study characterizes the visual language that \ST currently uses to communicate findings, thereby revealing opportunities for future visualization research to study exploratory workflows, interactive systems, and user-centered design practices beyond the static page.

\section{Conclusion}

This study characterized how \ST figures visually encode biological data. Using a \emph{What--Why--How} coding framework grounded in Munzner's nested model, we systematically characterized 148 papers, 1,824 figure panels, and the interaction affordances of 32 dedicated tools. Spatial grounding is preserved for tasks concerning the identity of biological entities and location, but recedes as reasoning moves toward functional, relational, or comparative analysis, where abstract idioms such as bar plots, chord diagrams, and UMAP embeddings dominate. Interactive tools show the same imbalance, supporting exploration and filtering far more consistently than the elaboration needed to move between an abstract summary and its spatial source. As \ST data grow in scale and complexity, we expect these gaps to widen rather than close on their own, underscoring an open and giving rise to exciting challenges and opportunities for biology and visualization researchers to explore together.

\section*{Acknowledgments}
The authors acknowledge support from the European Union’s Horizon Europe research and innovation programme through Marie Skłodowska-Curie Grant Agreement No. 101169349, as well as from the U.S. National Institutes of Health (U24 CA268108).

\bibliography{references}
\bibliographystyle{styles/abbrv-doi}

\raggedbottom

\newcommand{\compactbio}{%
  \vskip -2.5\baselineskip plus -1fil
}
\compactbio
\begin{IEEEbiographynophoto}{Denisse Chacón-Ramírez}
is a PhD student in the Visual Data Science Lab at Johannes Kepler University (JKU) Linz. Her research focuses on the visualization of single-cell and spatial biological data.
\end{IEEEbiographynophoto}

\compactbio
\begin{IEEEbiographynophoto}{Mark S. Keller}
is a postdoctoral researcher at Harvard Medical School (HMS) in the HIDIVE lab. His research focuses on interactive visualization tools for spatial and single-cell biomedical data.
\end{IEEEbiographynophoto}

\compactbio
\begin{IEEEbiographynophoto}{Eric Mörth}
is a Postdoctoral Research Fellow at Harvard Medical School. His work focuses on computational and visual analytics methods for large-scale 2D and 3D biological imaging data.
\end{IEEEbiographynophoto}

\compactbio
\begin{IEEEbiographynophoto}{Nils Gehlenborg, PhD}
is an Associate Professor of Biomedical Informatics at Harvard Medical School. His research develops visual interfaces and computational techniques that help scientists, clinicians, and patients interact with biomedical data.

\end{IEEEbiographynophoto}

\compactbio
\begin{IEEEbiographynophoto}{Marc Streit}
is a Full Professor of Visual Data Science at Johannes Kepler University Linz, Austria. His research interests include visualization, visual analytics, and explainable AI.
\end{IEEEbiographynophoto}

\compactbio
\begin{IEEEbiographynophoto}{Andreas Hinterreiter}
is a postdoctoral researcher and University Assistant in the Visual Data Science Lab at Johannes Kepler University (JKU) Linz. His research focuses on dimensionality reduction and explainable AI.
\end{IEEEbiographynophoto}

\end{document}